\documentclass[aps,prb,twocolumn,showpacs,floatfix,superscriptaddress]{revtex4-2}
\usepackage{graphicx,epsfig,amssymb}
\usepackage{color}
\usepackage{amsmath}
\usepackage{tikz}
\usepackage[utf8]{inputenc}
\usepackage{etoolbox}
\usepackage{booktabs}
\usepackage[breaklinks]{hyperref}
\hypersetup{breaklinks=true}
\hypersetup{hidelinks}
\usepackage{tabularx}
\usepackage{placeins}
\usepackage[utf8]{inputenc}
\usepackage{float}%\usepackage[style=phys,articletitle=false,biblabel=brackets,chaptertitle=false,pageranges=false}
\begin{document}

\raggedbottom

%\hypersetup{breaklinks=true}
%\urlstyle{same}

% Title of your paper
\title{Interface-Induced Ferromagnetism  in lateral $\mathrm{NiBr_{2}}$ and $\mathrm{NiCl_{2}}$ Heterostructure}

\author{Ilkay Ozdemir}
\affiliation{Physics Department, Adnan Menderes University, 09100 Aydin, T\"urkiye}
\affiliation{Department of Physics, University of Antwerp, Groenenborgerlaan 171, B-2020 Antwerp, Belgium}

\author{Mahsa Seyedmohammadzadeh}
\affiliation{Physics Department, Adnan Menderes University, 09100 Aydin, T\"urkiye}

\author{Yusuf Y\"{u}ksel}
\affiliation{Physics Department, Faculty of Science, Tinaztepe Campus, Dokuz Eylul University, 35390 Izmir, T\"urkiye}

\author{Olcay Uzengi Akt\"urk}
\affiliation{Department of Electrical Electronic Engeneering , Adnan Menderes University, Ayd{\i}n 09010, T\"urkiye}

\author{\"{U}mit Ak{\i}nc{\i}}
\affiliation{Physics Department, Faculty of Science, Tinaztepe Campus, Dokuz Eylul University, 35390 Izmir, T\"urkiye.}

 \author{Milorad V. Milo\v{s}evi\'{c}}
 \affiliation{Department of Physics, University of Antwerp,Groenenborgerlaan 171, B-2020,  Antwerp, Belgium}
\author{Seth Ariel Tongay}
\affiliation{Materials Science and Engineering, School for Engineering of Matter, Transport and Energy, Arizona State University, Tempe, Arizona, AZ 85287 USA}

\author{Johannes V. Barth}
\affiliation{Physics Department E20, Technical University of Munich (TUM), James Franck Strasse1, 85748, Garching (Germany)}

\author{Ethem Akt\"urk}
\email{ethem.akturk@adu.edu.tr}
\affiliation{Physics Department, Adnan Menderes University, 09100 Aydin, T\"urkiye}
\affiliation{Nanotechnology Application and Research Center, Adnan Menderes University, Ayd{\i}n 09010, T\"urkiye}

\date{\today}

%\keywords{Imaginary dielectric function, DFT, GW, BSE}

% Abstract
\begin{abstract}
Magnetic skyrmions are promising candidates for future information storing and processing devices. There are different routes for stabilizing the skyrmions. Understanding the interplay mechanism between different scenarios of skyrmion formation is one key factor that can reveal new paths for controlling skyrmion phases. Inspired by the flexibility of two-dimensional materials that offer an exciting playground for manipulating spin textures, we conducted \textit{ab initio} simulations and utilized four-state spin framework to determine magnetic parameters of lateral heterostructure formed by $\mathrm{NiBr_{2}}$ and $\mathrm{NiCl_{2}}$ Monolayers. The obtained spin interaction parameters are utilized to determine the skyrmionic phases via Monte Carlo simulation. Monte Carlo simulation results suggest three distinct phase transition mechanisms exist in the present system. Namely, examination of heat capacity versus temperature curve obtained from average anisotropic exchange energy yields a phase transition between paramagnetic and spin-spiral states at $5$K for pristine $\mathrm{NiBr_{2}}$, a paramagnetic-mixed skyrmion state transition occurs at $17$K for pristine $\mathrm{NiCl_{2}}$, and a high-temperature ferromagnetic-paramagnetic transition at $T=80$K is observed for the heterostructure region, indicating that some kind of intrinsic ferromagnetism may originate at the interface of pristine Janus structures.  \end{abstract}

% Make the title
\maketitle

% Main body of the paper
\section{Introduction}

In most magnetic materials, the spin exchange interaction leads to parallel or anti-parallel configuration of the magnetic moments. However, antisymmetric spin exchange described by the Dzyaloshinskii-Moriya interaction (DMI)~\cite{dzyaloshinsky1958thermodynamic,PhysRev.120.91} demonstrated that the lack of inversion symmetry and strong spin-orbit coupling can result in noncollinear spin arrangements, which, under specific conditions, can create long-lifetime topologically protected textures formed by swirling local magnetic-moment patterns, known as magnetic skyrmions (SkX)~\cite{bode2007chiral,ferriani2008atomic,heinze2011spontaneous,boulle2016room,muh}.\\
 Although utilizing the DMI or tuning its strength is the most recognized route to skyrmion formation~\cite{dai2023electric,kuznetsov2023effective,Ba2021,muh,janoschek2013fluctuation,mishra2023quantum}, other mechanisms exist that, individually or through their collective interaction, can facilitate the formation of skyrmions~\cite{nagaosa2013topological,PhysRevB.102.060402,https://doi.org/10.1002/adma.202107779,janson2014}. For instance, long-ranged magnetic dipolar interactions can give rise to magnetic SkX~\cite{TAKAO19831009,garel1982phase}. Also, stress-induced magnetic anisotropy can be utilized for creating and annihilating SkX in chiral magnets ~\cite{nii2015uniaxial}. Furthermore, it has been shown in frustrated systems such as centrosymmetric lattices, the intertwin between magnetic frustration and anisotropies of short-range symmetric exchange can lead to a unique chiral interaction that leads to the emergence of SkX~\cite{Amoroso2020,https://doi.org/10.1002/adma.201600889,Khanh2020,doi:10.1073/pnas.1118496109,PhysRevLett.120.077202}. On top of all, the isotropic higher-order interactions, which are usually investigated in the framework of four-spin four-site interactions  ~\cite{PhysRevB.102.014457,PhysRevB.84.224429,C2DT31662E}, are another route for stabilizing SkX~\cite{heinze2011spontaneous,PhysRevB.101.024418,PhysRevB.98.060413,PhysRevMaterials.7.054006}.\\
 Understanding the interplay between different routes of skyrmion formation is critical in advancing our comprehension of creating and controlling skyrmion, which maintains essential promise for integration into future technological applications. Two-dimensional (2D) materials are promising for examining such phenomena as they provide various geometrical designs to fix or break the symmetry. Notable examples are Janus 2D materials, where the inversion symmetry is inherently broken~\cite{das2023skyrmion,dou2022theoretical,han2023strain,das2023skyrmion,
 li2023topological,liang2020very,xu2020topological,zhang2020emergence,yagmurcukardes2020quantum,xiao2024strain}, and van der Waals heterostructures (vdWHs), where Rashba spin-orbit coupling can induce DMI~\cite{doi:10.1021/acs.nanolett.7b04722,Sun2020,https://doi.org/10.1002/adfm.202104452,PhysRevB.107.184439,huang2024magnetic}.\\ 

In light of recent progress in fabricating 2D magnetic heterostructures~\cite{doi:10.1021/acs.chemmater.3c00172} and considering the significant role of SkX in new technological frontiers~\cite{shen2023programmable,doi:10.1021/acs.nanolett.7b04722}, in this paper, we have focused our investigation on lateral heterostructure formed by $\mathrm{NiBr_{2}}$ and $\mathrm{NiCl_{2}}$. We employed an ab initio analysis to investigate the mechanical and electronic properties of the considered systems. The four-state energy mapping method~\cite{C2DT31662E,vsabani2020ab} is utilized to obtain magnetic parameters, and the results are passed as input to the Monte Carlo simulation code to determine SkX phases. We identified three distinct phase transition scenarios. A paramagnetic to spin-spiral states for $\mathrm{NiBr_{2}}$ , a paramagnetic-mixed skyrmion state transition for $\mathrm{NiCl_{2}}$, and a high-temperature ferromagnetic-paramagnetic transition at T = 80K for the heterostructure region.      

\begin{figure}[!ht]
\centering
\includegraphics[width=1\columnwidth]{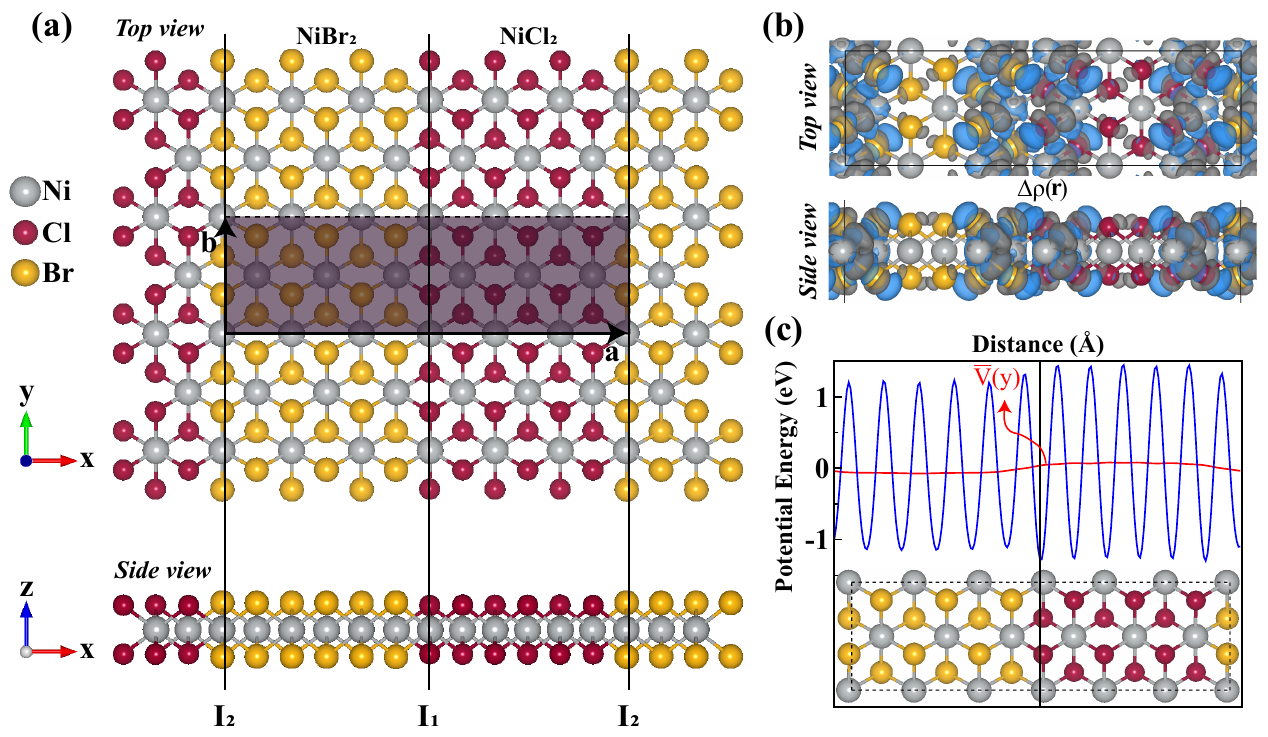}
\caption{\textbf{(a)} Top and side views of lateral heterostructure formed by $\mathrm{NiBr_{2}}$ and $\mathrm{NiCl_{2}}$. \textbf{(b)} Charge density difference between isolated layers and lateral heterostructure. \textbf{(c)} Electrostatic potential across the heterostructure.}
\label{Fig1}
\end{figure}

\section{Computational Details}

We performed ab initio calculations based on density functional theory (DFT)~\cite{PhysRev.140.A1133}  within the framework of projector-augmented wave potentials~\cite{PhysRevB.50.17953,PhysRevB.50.17953} and generalized gradient approximation of Perdew, Burke, and Ernzerhof (PBE)~\cite{PhysRevLett.77.3865} as implemented in VASP (Vienna Ab initio Simulation Package) software~\cite{PhysRevB.47.558,KRESSE199615}.\\
 As illustrated in Fig.~S1~\cite{ESI} and Fig.~\ref{Fig1}(a), we used a fully relaxed orthogonal cell to create lateral heterostructures. Then, we performed geometry optimization using the following parameters: a $\mathrm{\Gamma}$-centered $2 \times 8 \times 1$ employed for Brillouin zone integration, an energy cutoff of 450 eV was chosen to represent the plane-wave basis, the electronic convergence criterion was set to $1.0 \times 10^{-5}$ eV, a vacuum of 20 $\mathrm{\AA}$ is considered above the layers to eliminate the interaction between periodic images and a Gaussian smearing method was applied with a smearing parameter of 0.01 eV. The ionic relaxation algorithm was based on the conjugate gradient method, and both atomic positions and cell shape were fully relaxed. The ionic convergence criterion was set to $1 \times 10^{-3}$ eV/$\mathrm{\AA}$. In all calculations, spin-orbit coupling is considered to account for magnetic properties. Furthermore, to effectively handle the substantial on-site interactions characteristic of Ni-3d electrons, PBE+U within Dudarev’s approach~\cite{PhysRevB.57.1505} is employed.\\
 To confirm the dynamic stability of the relaxed structures, we employed density functional perturbation theory~\cite{RevModPhys.73.515} to calculate second-order force constants using $1 \times 2 \times 1$ supercell and $2 \times 4 \times 1$ k-point mesh. Phonon frequencies are calculated using the Phonopy code\cite{TOGO20151}. In order to get a better insight into electronic properties, we used the Heyd-Scuseria-Ernzerhof hybrid functional (HSE06)~\cite{krukau2006influence,heyd2003hybrid}.\\
 We implemented the four-state energy mapping method to calculate the exchange coupling interactions tensor $\mathrm{J_{ij}}$. For instance, to obtain the $\mathrm{J_{xz}}$ term for spins located at lattice points with indices $\mathrm{i=1}$ and $\mathrm{j=2}$, the required configurations are as follows:
\begin{itemize}
\item State 1: $\vec{S}_{1}=(+S,0,0)$; $\vec{S}_{2}=(0,0,+S)$
\item State 2: $\vec{S}_{1}=(+S,0,0)$; $\vec{S}_{2}=(0,0,-S)$
\item State 3: $\vec{S}_{1}=(-S,0,0)$; $\vec{S}_{2}=(0,0,+S)$
\item State 4: $\vec{S}_{1}=(-S,0,0)$; $\vec{S}_{2}=(0,0,-S)$
\end{itemize}
Except for spins with indices $\mathrm{i=1}$ and $\mathrm{j=2}$, all other spins are in the $(0,+S,0)$ or $(0,-S,0)$ state. If the energies obtained from calculating these four states are denoted as $\mathrm{E_1, E_2, E_3, E_4}$ respectively, the $\mathrm{J_{xz}}$ term for spins located at lattice points with indices $\mathrm{i=1}$ and $\mathrm{j=2}$, $\mathrm{J_{12}^{xz}}$, can be determined by substituting into the equation~\cite{C2DT31662E,vsabani2020ab}:
\begin{equation} \label{equ:1}
\mathrm{J_{12}^{xz}=\frac{E_1+E_4-E_2-E_3}{4S^2}}
\end{equation}

\section{Results}
\begin{table}[!ht]
    \centering
 \caption{Calculated lattice constants in the x-direction (\(\mathrm{a}\)) and y-direction (\(\mathrm{b}\)),band gaps obtained using DFT \(\mathrm{E_{g}^{DFT}}\) and HSE06 \(\mathrm{E_{g}^{HSE}}\).}
    \begin{tabularx}{0.5\textwidth}{XXXX}
    \hline \hline \\[0.1 em]
    \(\mathrm{a}\) (\(\mathrm{\AA}\)) & \(\mathrm{b}\) (\(\mathrm{\AA}\)) & \(\mathrm{E_{g}^{DFT}}\) (\(\mathrm{eV}\)) & \(\mathrm{E_{g}^{HSE}}\) (\(\mathrm{eV}\)) \\[0.5 em]
    \hline \\[0.5 em]
    21.53 & 6.22 & 1.09 & 4.15  \\[0.5 em]
    \hline  \hline
    \end{tabularx}
    \label{table:additional_data}
\end{table}

We presented details of the relaxed structure in Table 1. The obtained lattice constants in the x and y directions are  6.22 $(\mathrm{\AA})$ and 21.53 $(\mathrm{\AA})$, respectively. There is an increasing trend in the distances between Ni atoms when moving from the center of the heterostructure in the x-direction to the left and right sides of the interface Ni atoms. This behavior is asymmetric in the $\mathrm{NiBr_{2}}$ and $\mathrm{NiCl_{2}}$ sides of the heterostructure as the distances between Ni atoms on the $\mathrm{NiBr_{2}}$ side are larger. The calculated cohesive energy of the considered heterostructure is 3.00 ($\mathrm{eV/atom}$), which indicates the chemical stability of the system.   Besides,  according to the calculated phonon dispersion curves shown in Fig.~S2~\cite{ESI}, the structure possesses good dynamic stability. \\

\begin{figure*}[!ht]
\centering
\includegraphics[width=2\columnwidth]{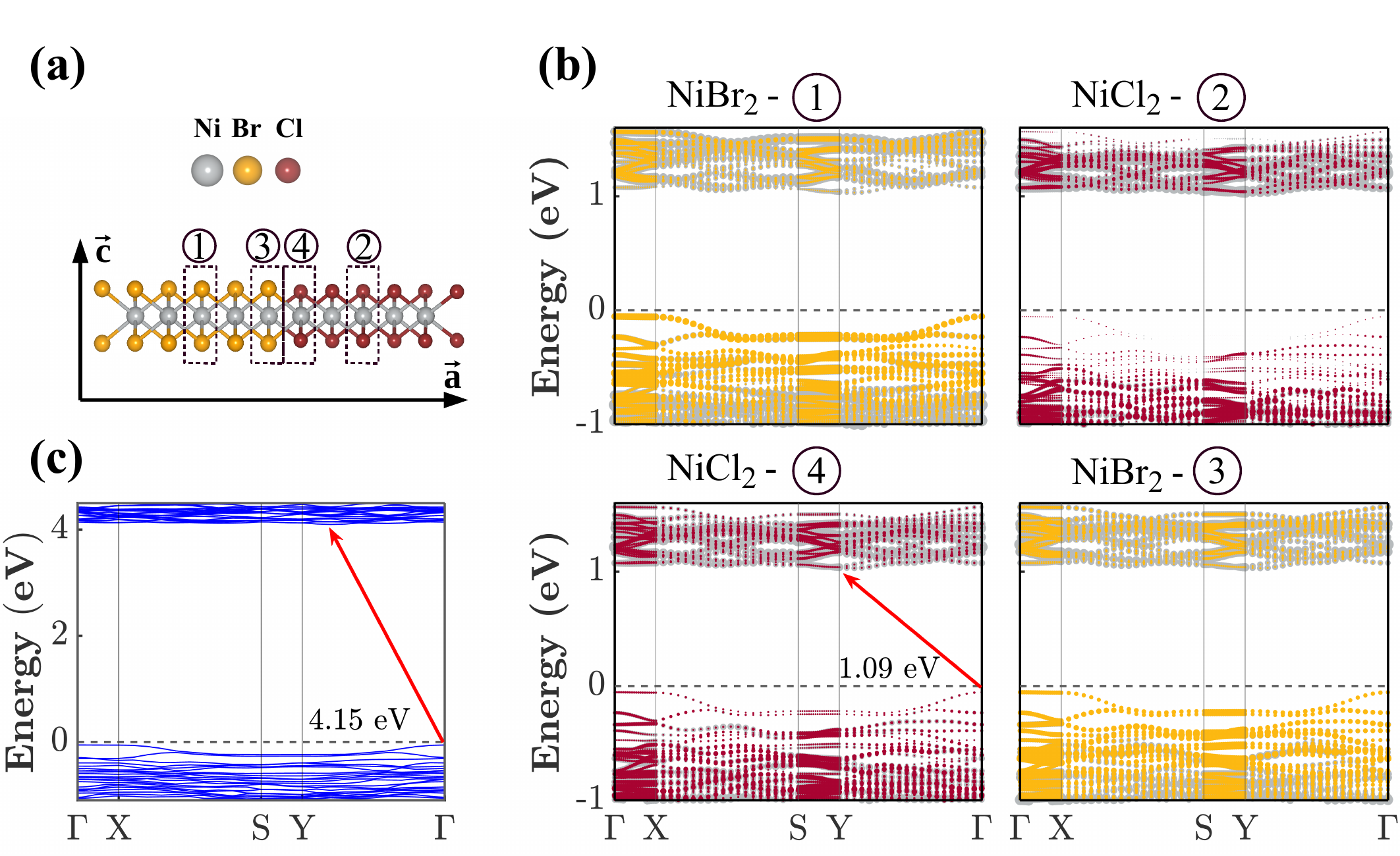}
\caption{$\mathbf{(a)}$ $\mathrm{NiBr_{2}}$/$\mathrm{NiCl_{2}}$ heterostructure. $\mathbf{(b)}$ DFT band structures projected onto $\mathrm{Br-Ni-Br}$ and $\mathrm{Cl-Ni-C}$, as labeled in $\mathbf{(a)}$. Zero of energy is set to the Fermi level. $\mathbf{(c)}$ HSE bandstructure}
\label{Fig2}
\end{figure*}

 Our DFT calculation predicts an indirect band gap of 1.09 (eV), while HSE gives a band gap of 4.15 (eV), which is a typical behavior of systems with strongly correlated electrons. Fig.~\ref{Fig2} (b) depicts the contribution of atoms from different regions of the heterostructure on electronic band structure. The influence of the $\mathrm{NiBr_{2}}$ side on the top valence band is predominant, and the contribution of non-interface Br atoms to the system is significantly enhanced. On the other hand, the non-interface Ni atoms of the $\mathrm{NiCl_{2}}$ side have the most contribution in the first conduction band. These results can be attributed to the higher electron negativity of Br atoms and the asymmetry in the distribution of neighboring atoms. As illustrated in Fig.~\ref{Fig1}(c), the average electrostatic potential in $\mathrm{NiBr_{2}}$ side is slightly larger than $\mathrm{NiCl_{2}}$ side. This result align with the contribution of atoms in the band structure and show how the non-interface regions are distinguished.   

 Following the examination of electronic properties, we now shift our focus to the investigation of magnetic properties. To this aim, we first calculated $\mathbf{J}_{ij}$ tensor and the isotropic exchange interaction constant by using the four-states method \cite{Amoroso2020}. For quantitative details, refer to the supplementary materials~\cite{ESI}. Exchange interaction parameters for the pristine $\mathrm{NiBr_{2}}$ and $\mathrm{NiCl_{2}}$ monolayers have been extracted from Ref. \cite{Amoroso2020}.
We applied Monte Carlo (MC) method based on the standard Metropolis algorithm \cite{landau2014guide} with Marsaglia spin-update scheme \cite{marsaglia1972choosing} to simulate the magnetic properties of $\mathrm{NiX_{2}+NiY_{2}(X=Br,Y=Cl)}$ hetero-structure with the following atomistic spin Hamiltonian
%\begin{equation}\label{eq1}
%	\mathcal{H}=\frac{1}{2}\sum_{i\neq %j}\mathbf{s}_{i}\mathbf{J}_{ij}\mathbf{s}_{j}+\sum_{i}\mathbf{s}_{i}\mathbf{A}_{i}s_{i}-\mathbf{B}\sum_{i%}\mathbf{s}_{i},
%\end{equation}
\begin{equation}\label{eq1_mc}
	\mathcal{H}=\frac{1}{2}\sum_{i\neq j}\mathbf{s}_{i}\mathbf{J}_{ij}\mathbf{s}_{j}-\mathbf{B}\sum_{i}\mathbf{s}_{i},
\end{equation}
where $\mathbf{s}_{i}$ is a classical spin vector with unit magnitude, $\mathbf{J}_{ij}$ is the exchange interaction tensor, and $\mathbf{B}=B\hat{k}$ represents the applied magnetic field.\

 To investigate the magnetic properties, the overall triangular lattice with dimensions $(L_{x}=3L,L_{y}=L)$ is divided into three parts as shown in Fig.~S4~\cite{ESI} where the left-most and the right-most parts are respectively partitioned for pristine $\mathrm{NiBr_{2}}$ and $\mathrm{NiCl_{2}}$ mono-layers whereas the intermediate region is defined as the lateral heterostructure with the half-width $\sigma$. For the calculation of the magnetic properties, we imposed periodic boundary conditions in each direction for a lattice with $L=60$, and we consider $2.5\times10^{5}$ Monte Carlo steps per lattice site. We monitored the relevant quantities after discarding the initial $10\%$ of the overall data  for the thermalization. Also, in order to reduce the statistical errors in temperature dependent properties such as the magnetization and heat capacity, we performed sample averaging over 20 independent realizations at each temperature step. Initially, we start from a high temperature state where all spin directions are randomized, and we gradually reduce the temperature down to $T=0.01$K.    
The magnetic field is scaled as $h=B/|J_{1}^{iso}|$ where $J_{1}^{iso}$ is the isotropic nearest-neighbor exchange constant of pristine $\mathrm{NiCl_2}$ \cite{Amoroso2020}.

\begin{figure*}[!ht]
    \centering
    % First figure with label (a)
    \begin{tikzpicture}
        \node[anchor=south west,inner sep=0] (image) at (0,0) {\includegraphics[width=0.3\textwidth]{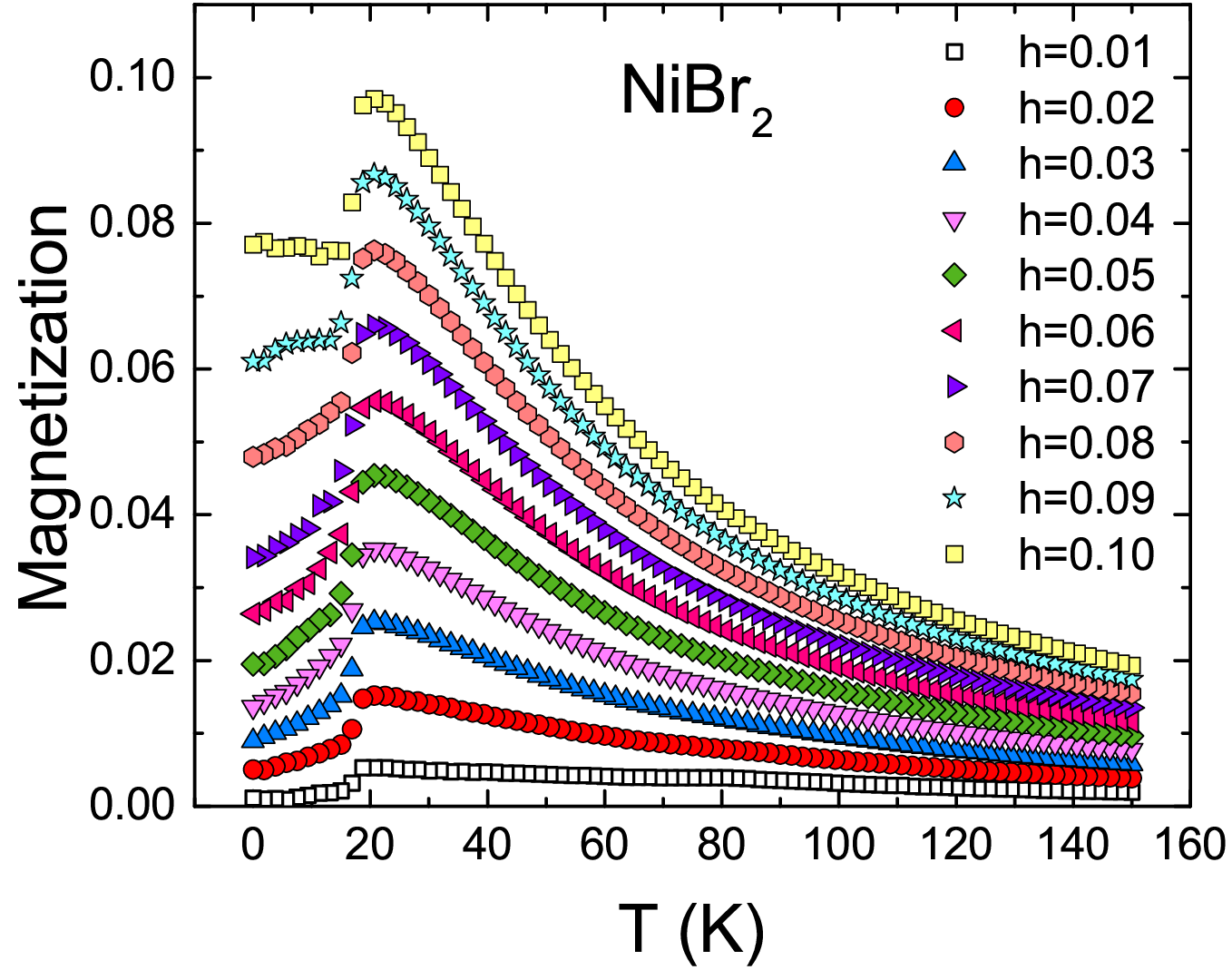}};
        \node[anchor=north west] at ([xshift=10pt, yshift=16pt]image.north west) {\bfseries\large (a)};
    \end{tikzpicture}
    \hfill % Fills the space between the figures
    % Second figure with label (b)
    \begin{tikzpicture}
        \node[anchor=south west,inner sep=0] (image) at (0,0) {\includegraphics[width=0.295\textwidth]{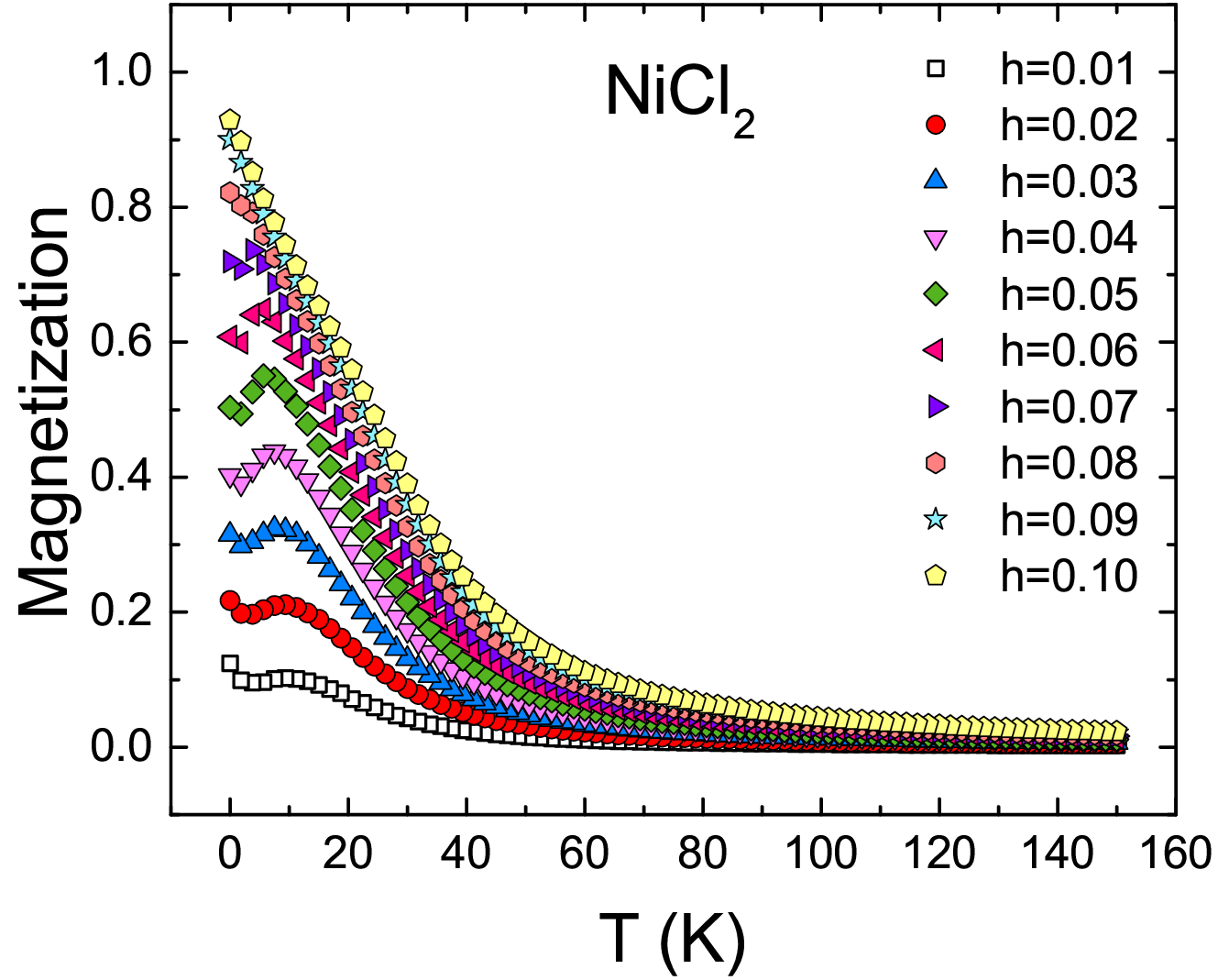}};
        \node[anchor=north west] at ([xshift=10pt, yshift=16pt]image.north west) {\bfseries\large (b)};
    \end{tikzpicture}
    \hfill % Fills the space between the figures
    % Third figure with label (c)
    \begin{tikzpicture}
        \node[anchor=south west,inner sep=0] (image) at (0,0) {\includegraphics[width=0.3\textwidth]{fig3c.eps}};
        \node[anchor=north west] at ([xshift=10pt, yshift=16pt]image.north west) {\bfseries\large (c)};
    \end{tikzpicture}
    \caption{Temperature dependence of magnetization obtained for (a) pristine $\mathrm{NiBr_{2}}$, (b) heterostructure region, (c) pristine $\mathrm{NiCl_{2}}$ with the corresponding applied magnetic field values.}
   \label{fig3}
\end{figure*}

 Variation of magnetization of the system as a function of temperature is given in Fig.~\ref{fig3} for pristine $\mathrm{NiBr_{2}}$ and $\mathrm{NiCl_{2}}$ monolayers, as well as for the hetero-structure region with some selected values of external magnetic field. From Fig. \ref{fig3} (a), it is clear that pristine $\mathrm{NiBr_{2}}$ monolayer magnetization remains zero for small values of $h$ whereas with magnetic field enhancement, saturation magnetization tend to saturate at a small non-zero value which shows that a phase transition from paramagnetic to spin-spiral (SS) state may take place at a critical temperature. The location of this transition temperature does not depend on $h$. For $\mathrm{NiCl_{2}}$ pristine monolayer (Fig.~\ref{fig3} (b)), we see that a well saturated magnetization state is achieved with increasing $h$. On the other hand, for $h\leq 0.1$, behavior of magnetization at low temperature region shows that some skyrmion-like vortices may emerge in the pristine  $\mathrm{NiCl_{2}}$ (Fig.~\ref{fig3} (c)). Furthermore, for the hetero-structure region (Fig. \ref{fig3} (b)), a phase transition between paramagnetic and ferromagnetic states takes place due to enhanced diagonal exchange energy. Note that the external field is sufficiently weak to destruct this phase transition behavior.\

\begin{figure}[!ht]
    \centering
	\includegraphics[width=\columnwidth]{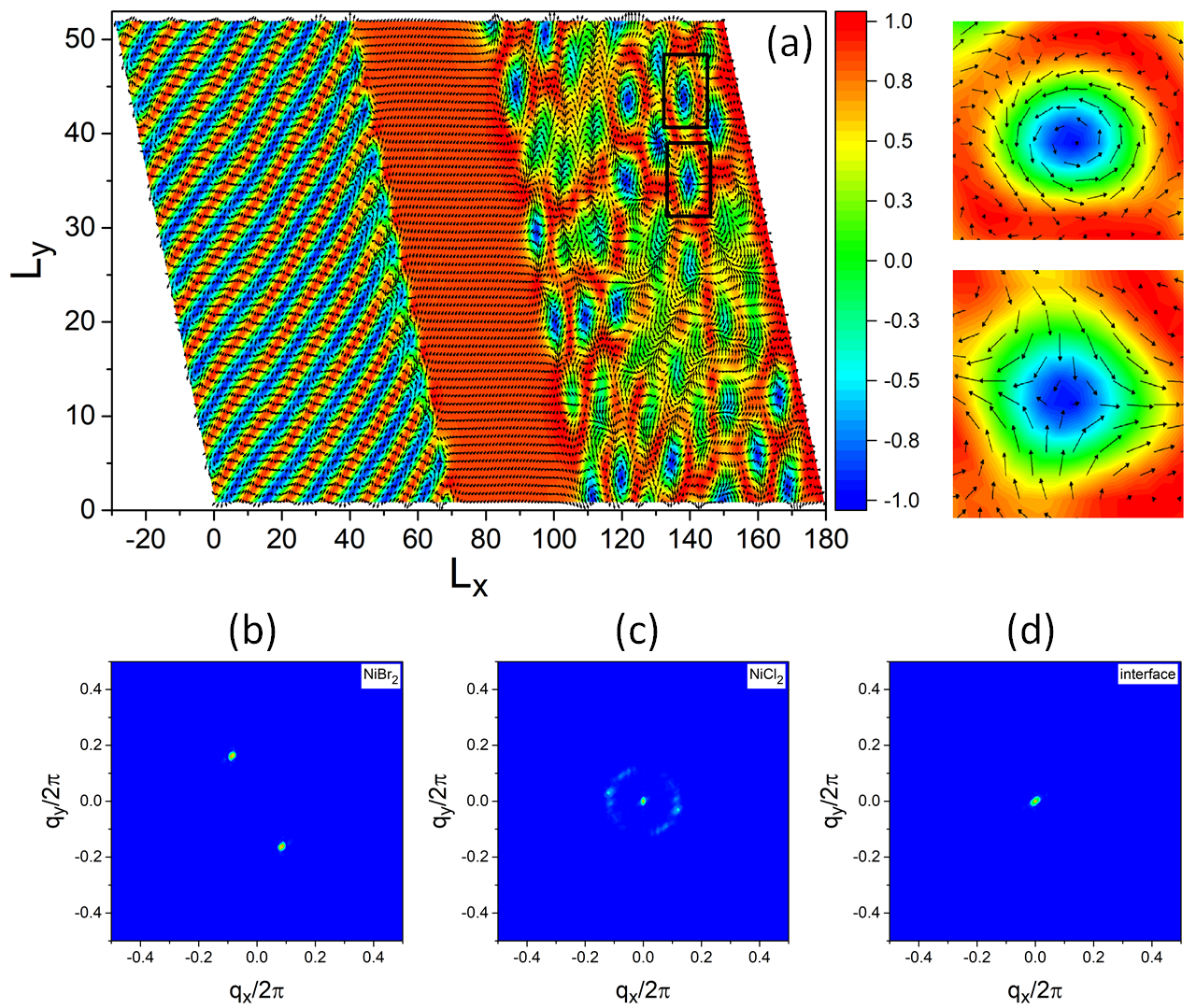}
	\caption{(a) Real-space spin configurations of the overall lattice obtained for $h=0.04$. The magnified Bloch-type skyrmion and anti-skyrmion formations are marked with the rectangles. (b)-(d) Structure factor plots in respective order for pristine $\mathrm{NiBr_{2}}$ and $\mathrm{NiCl_{2}}$ monolayers, as well as hetero-structure regime.}\label{fig4}
\end{figure}

\begin{figure}[!ht]
    \centering
	\includegraphics[width=1\columnwidth]{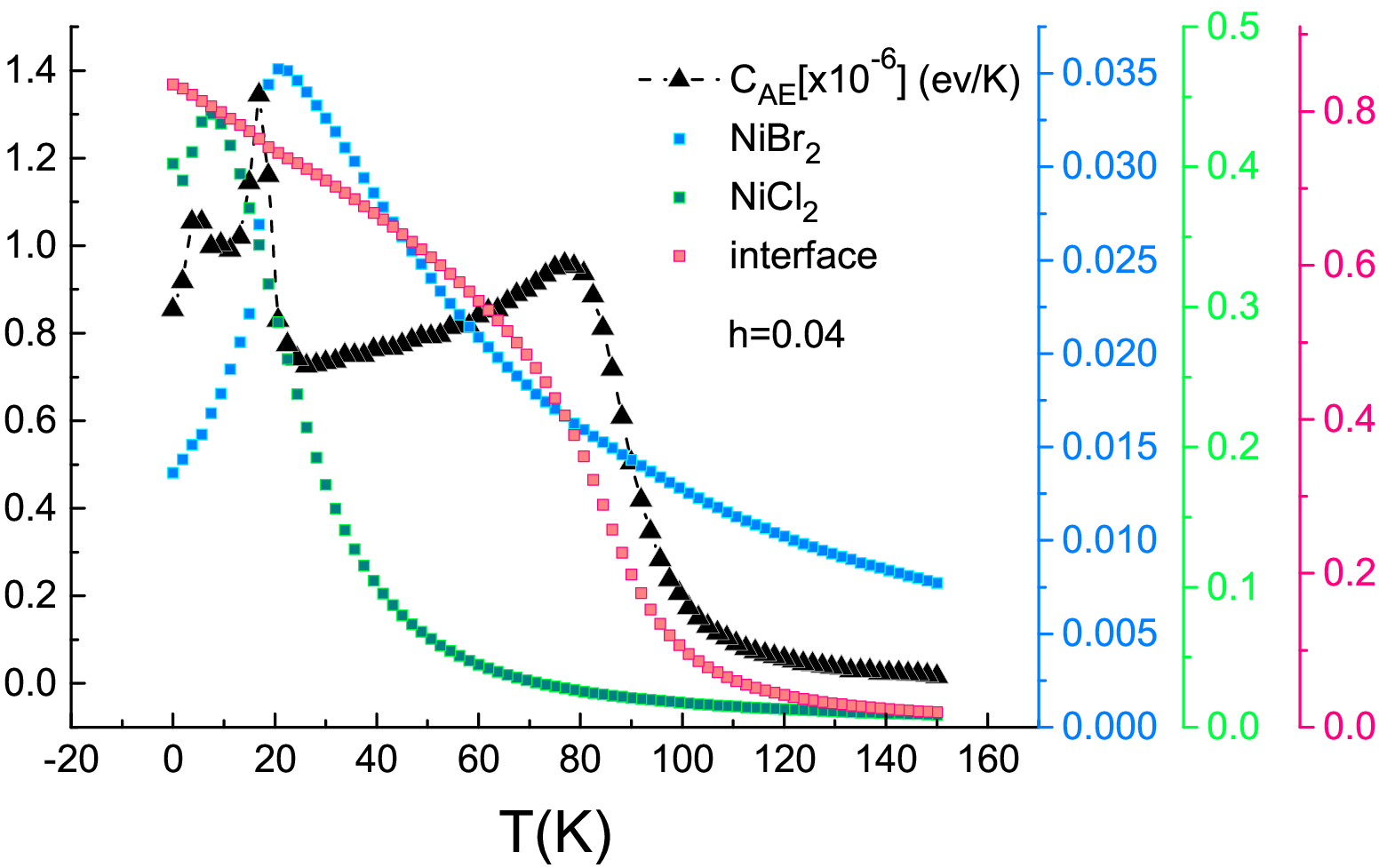}
	\caption{Temperature dependent  magnetization curves, as well as the total magnetic specific heat obtained from anisotropic exchange energy curve for $h=0.04$.}\label{fig5}
\end{figure}

Real-time spin configurations obtained at the ground state are depicted in Fig. \ref{fig4} (a) corresponding to $h=0.04$. It is clear that pristine $\mathrm{NiBr_{2}}$ exhibits a spin-spiral state, and according to our simulations, spin-spiral state survives for field values up to $h=0.1$. For $\mathrm{NiCl_{2}}$, we observe a mixed state composed of both Bloch type skyrmion vortices and anti-skyrmion particles (see the magnified plots in Fig. \ref{fig4} (a). As the magnetic field gradually increases, vortex domains disappear and a field polarized (FP) state emerges. On the other hand, the intermediate (heterostructure) region exhibits a ferromagnetic state with full alignment of dipole moments along the applied field direction.  We also calculate the static structure factor using the relation:

\begin{equation}\label{eq2_mc}
\begin{split}
	S(\mathbf{q})=\frac{1}{N}\Biggl\langle &\left|\sum_{\mathbf{r}}S^{(x)}_{\mathbf{r}}\exp(-i\mathbf{q}\cdot\mathbf{r})\right|^{2} \\
	&+ \left|\sum_{\mathbf{r}}S^{(y)}_{\mathbf{r}}\exp(-i\mathbf{q}\cdot\mathbf{r})\right|^{2} \\
	&+ \left|\sum_{\mathbf{r}}S^{(z)}_{\mathbf{r}}\exp(-i\mathbf{q}\cdot\mathbf{r})\right|^{2} \Biggr\rangle, 
\end{split}
\end{equation}

As shown in lower panels of Fig. \ref{fig4}, the intensity of the structure factor exhibits two and one prominent peaks for SS (Fig. \ref{fig4} (b)) and FP (Fig. \ref{fig4} (d)) phases, respectively. Whereas for the hetero-structure region, we observe one central peak along with several perimeter spots characterizing the mixed vortex pattern scenario illustrated in Fig.~\ref{fig4}.

 We found that there are three distinct and particular phase transition mechanisms which can be attributed to three partitions considered in the present hetero-structure system. In order to estimate the location of the transition temperatures, we examine the temperature variation of total heat capacity obtained from average anisotropic exchange energy for $h=0.04$. The results are displayed in Fig. \ref{fig5} where we also plot the magnetization of each partition for comparison. As shown in the figure, heat capacity exhibits three successive peaks. The low temperature peak emerging around $5$K is associated to the paramagnetic-mixed phase (skyrmion+anti-skyrmion) transition in pristine $\mathrm{NiCl_{2}}$. The second peak observed at $17$K corresponds to paramagnetic-spin spiral state phase transition in pristine $\mathrm{NiBl_{2}}$ whereas the high temperature cusp at $80$K accompanies the ferromagentic-paramagnetic phase transition in the hetero-structure region.  As a final remark, we emphasize that the half-width ($\sigma$) of the hetero-structure region does not have any noticeable impact on the presented results, and the main finding will be robust both qualitatively and quantitatively for some other selected $\sigma$ values.

 \section{CONCLUSION}
 We first examined the chemical and dynamical stability of the lateral heterostructure formed by $\mathrm{NiBr_{2}}$ and $\mathrm{NiCl_{2}}$ monolayers. Then, utilizing DFT calculations and MC simulations, we systematically investigated its electronic and magnetic properties.
 Decomposing the band structure to the contribution of atoms shows three regions with distinguished electronic properties. The non-interface Br and Ni atoms dominate the top valence band. In contrast, non-interface Ni atoms mainly form the first conduction band. The contribution of the interface atoms from both sides of the heterostructure to the first conduction and valence band are almost similar.
We employed the four-state method for obtaining the exchange interaction of interface atoms and then performed MC simulation to investigate the magnetic properties of the system.  Regarding the magnetic properties of the simulated lateral hetero-structure, we have obtained three distinct phase transition characteristics obtained by examining the temperature dependent magnetization, as well as magnetic specific heat capacity curves. Namely, pristine $\mathrm{NiBr_{2}}$ and $\mathrm{NiCl_{2}}$ lattices respectively exhibit spin-spiral and skyrmion/anti-skyrmion mixed phases in the ground state whereas the interfacial hetero-structure exhibits ferromagnetism with Curie point located around  80 K temperature. These results suggest that the hetero-structure interface  may exhibit intrinsic magnetic phenomena which can be very different from the corresponding pristine Janus counterparts.

\section{ACKNOWLEDGMENTS}

This work is supported by The Scientific and Technological Research Council of Turkey (T\"{U}BITAK) under Grant No. 122F142. The computational resources are provided by T\"UBITAK ULAKBIM, High Performance and Grid Computing Center (TR-Grid e-Infrastructure) and the Leibniz Supercomputing Centre.  E. A. acknowledges the Alexander von Humboldt Foundation for a Research Fellowship for Experienced Researchers.

\FloatBarrier
\bibliography{paper}

\begin{thebibliography}{60}%
\makeatletter
\providecommand \@ifxundefined [1]{%
 \@ifx{#1\undefined}
}%
\providecommand \@ifnum [1]{%
 \ifnum #1\expandafter \@firstoftwo
 \else \expandafter \@secondoftwo
 \fi
}%
\providecommand \@ifx [1]{%
 \ifx #1\expandafter \@firstoftwo
 \else \expandafter \@secondoftwo
 \fi
}%
\providecommand \natexlab [1]{#1}%
\providecommand \enquote  [1]{``#1''}%
\providecommand \bibnamefont  [1]{#1}%
\providecommand \bibfnamefont [1]{#1}%
\providecommand \citenamefont [1]{#1}%
\providecommand \href@noop [0]{\@secondoftwo}%
\providecommand \href [0]{\begingroup \@sanitize@url \@href}%
\providecommand \@href[1]{\@@startlink{#1}\@@href}%
\providecommand \@@href[1]{\endgroup#1\@@endlink}%
\providecommand \@sanitize@url [0]{\catcode `\\12\catcode `\$12\catcode
  `\&12\catcode `\#12\catcode `\^12\catcode `\_12\catcode `\%12\relax}%
\providecommand \@@startlink[1]{}%
\providecommand \@@endlink[0]{}%
\providecommand \url  [0]{\begingroup\@sanitize@url \@url }%
\providecommand \@url [1]{\endgroup\@href {#1}{\urlprefix }}%
\providecommand \urlprefix  [0]{URL }%
\providecommand \Eprint [0]{\href }%
\providecommand \doibase [0]{https://doi.org/}%
\providecommand \selectlanguage [0]{\@gobble}%
\providecommand \bibinfo  [0]{\@secondoftwo}%
\providecommand \bibfield  [0]{\@secondoftwo}%
\providecommand \translation [1]{[#1]}%
\providecommand \BibitemOpen [0]{}%
\providecommand \bibitemStop [0]{}%
\providecommand \bibitemNoStop [0]{.\EOS\space}%
\providecommand \EOS [0]{\spacefactor3000\relax}%
\providecommand \BibitemShut  [1]{\csname bibitem#1\endcsname}%
\let\auto@bib@innerbib\@empty
%</preamble>
\bibitem [{\citenamefont
  {Dzyaloshinsky}(1958)}]{dzyaloshinsky1958thermodynamic}%
  \BibitemOpen
  \bibfield  {author} {\bibinfo {author} {\bibfnamefont {I.}~\bibnamefont
  {Dzyaloshinsky}},\ }\bibfield  {title} {\bibinfo {title} {A thermodynamic
  theory of “weak” ferromagnetism of antiferromagnetics},\ }\href@noop {}
  {\bibfield  {journal} {\bibinfo  {journal} {Journal of physics and chemistry
  of solids}\ }\textbf {\bibinfo {volume} {4}},\ \bibinfo {pages} {241}
  (\bibinfo {year} {1958})}\BibitemShut {NoStop}%
\bibitem [{\citenamefont {Moriya}(1960)}]{PhysRev.120.91}%
  \BibitemOpen
  \bibfield  {author} {\bibinfo {author} {\bibfnamefont {T.}~\bibnamefont
  {Moriya}},\ }\bibfield  {title} {\bibinfo {title} {Anisotropic superexchange
  interaction and weak ferromagnetism},\ }\href
  {https://doi.org/10.1103/PhysRev.120.91} {\bibfield  {journal} {\bibinfo
  {journal} {Phys. Rev.}\ }\textbf {\bibinfo {volume} {120}},\ \bibinfo {pages}
  {91} (\bibinfo {year} {1960})}\BibitemShut {NoStop}%
\bibitem [{\citenamefont {Bode}\ \emph {et~al.}(2007)\citenamefont {Bode},
  \citenamefont {Heide}, \citenamefont {Von~Bergmann}, \citenamefont
  {Ferriani}, \citenamefont {Heinze}, \citenamefont {Bihlmayer}, \citenamefont
  {Kubetzka}, \citenamefont {Pietzsch}, \citenamefont {Bl{\"u}gel},\ and\
  \citenamefont {Wiesendanger}}]{bode2007chiral}%
  \BibitemOpen
  \bibfield  {author} {\bibinfo {author} {\bibfnamefont {M.}~\bibnamefont
  {Bode}}, \bibinfo {author} {\bibfnamefont {M.}~\bibnamefont {Heide}},
  \bibinfo {author} {\bibfnamefont {K.}~\bibnamefont {Von~Bergmann}}, \bibinfo
  {author} {\bibfnamefont {P.}~\bibnamefont {Ferriani}}, \bibinfo {author}
  {\bibfnamefont {S.}~\bibnamefont {Heinze}}, \bibinfo {author} {\bibfnamefont
  {G.}~\bibnamefont {Bihlmayer}}, \bibinfo {author} {\bibfnamefont
  {A.}~\bibnamefont {Kubetzka}}, \bibinfo {author} {\bibfnamefont
  {O.}~\bibnamefont {Pietzsch}}, \bibinfo {author} {\bibfnamefont
  {S.}~\bibnamefont {Bl{\"u}gel}},\ and\ \bibinfo {author} {\bibfnamefont
  {R.}~\bibnamefont {Wiesendanger}},\ }\bibfield  {title} {\bibinfo {title}
  {Chiral magnetic order at surfaces driven by inversion asymmetry},\
  }\href@noop {} {\bibfield  {journal} {\bibinfo  {journal} {nature}\ }\textbf
  {\bibinfo {volume} {447}},\ \bibinfo {pages} {190} (\bibinfo {year}
  {2007})}\BibitemShut {NoStop}%
\bibitem [{\citenamefont {Ferriani}\ \emph {et~al.}(2008)\citenamefont
  {Ferriani}, \citenamefont {Von~Bergmann}, \citenamefont {Vedmedenko},
  \citenamefont {Heinze}, \citenamefont {Bode}, \citenamefont {Heide},
  \citenamefont {Bihlmayer}, \citenamefont {Bl{\"u}gel},\ and\ \citenamefont
  {Wiesendanger}}]{ferriani2008atomic}%
  \BibitemOpen
  \bibfield  {author} {\bibinfo {author} {\bibfnamefont {P.}~\bibnamefont
  {Ferriani}}, \bibinfo {author} {\bibfnamefont {K.}~\bibnamefont
  {Von~Bergmann}}, \bibinfo {author} {\bibfnamefont {E.}~\bibnamefont
  {Vedmedenko}}, \bibinfo {author} {\bibfnamefont {S.}~\bibnamefont {Heinze}},
  \bibinfo {author} {\bibfnamefont {M.}~\bibnamefont {Bode}}, \bibinfo {author}
  {\bibfnamefont {M.}~\bibnamefont {Heide}}, \bibinfo {author} {\bibfnamefont
  {G.}~\bibnamefont {Bihlmayer}}, \bibinfo {author} {\bibfnamefont
  {S.}~\bibnamefont {Bl{\"u}gel}},\ and\ \bibinfo {author} {\bibfnamefont
  {R.}~\bibnamefont {Wiesendanger}},\ }\bibfield  {title} {\bibinfo {title}
  {Atomic-scale spin spiral with a unique rotational sense: Mn monolayer on w
  (001)},\ }\href@noop {} {\bibfield  {journal} {\bibinfo  {journal} {Physical
  review letters}\ }\textbf {\bibinfo {volume} {101}},\ \bibinfo {pages}
  {027201} (\bibinfo {year} {2008})}\BibitemShut {NoStop}%
\bibitem [{\citenamefont {Heinze}\ \emph {et~al.}(2011)\citenamefont {Heinze},
  \citenamefont {Von~Bergmann}, \citenamefont {Menzel}, \citenamefont {Brede},
  \citenamefont {Kubetzka}, \citenamefont {Wiesendanger}, \citenamefont
  {Bihlmayer},\ and\ \citenamefont {Bl{\"u}gel}}]{heinze2011spontaneous}%
  \BibitemOpen
  \bibfield  {author} {\bibinfo {author} {\bibfnamefont {S.}~\bibnamefont
  {Heinze}}, \bibinfo {author} {\bibfnamefont {K.}~\bibnamefont
  {Von~Bergmann}}, \bibinfo {author} {\bibfnamefont {M.}~\bibnamefont
  {Menzel}}, \bibinfo {author} {\bibfnamefont {J.}~\bibnamefont {Brede}},
  \bibinfo {author} {\bibfnamefont {A.}~\bibnamefont {Kubetzka}}, \bibinfo
  {author} {\bibfnamefont {R.}~\bibnamefont {Wiesendanger}}, \bibinfo {author}
  {\bibfnamefont {G.}~\bibnamefont {Bihlmayer}},\ and\ \bibinfo {author}
  {\bibfnamefont {S.}~\bibnamefont {Bl{\"u}gel}},\ }\bibfield  {title}
  {\bibinfo {title} {Spontaneous atomic-scale magnetic skyrmion lattice in two
  dimensions},\ }\href@noop {} {\bibfield  {journal} {\bibinfo  {journal}
  {nature physics}\ }\textbf {\bibinfo {volume} {7}},\ \bibinfo {pages} {713}
  (\bibinfo {year} {2011})}\BibitemShut {NoStop}%
\bibitem [{\citenamefont {Boulle}\ \emph {et~al.}(2016)\citenamefont {Boulle},
  \citenamefont {Vogel}, \citenamefont {Yang}, \citenamefont {Pizzini},
  \citenamefont {de~Souza~Chaves}, \citenamefont {Locatelli}, \citenamefont
  {Mente{\c{s}}}, \citenamefont {Sala}, \citenamefont {Buda-Prejbeanu},
  \citenamefont {Klein} \emph {et~al.}}]{boulle2016room}%
  \BibitemOpen
  \bibfield  {author} {\bibinfo {author} {\bibfnamefont {O.}~\bibnamefont
  {Boulle}}, \bibinfo {author} {\bibfnamefont {J.}~\bibnamefont {Vogel}},
  \bibinfo {author} {\bibfnamefont {H.}~\bibnamefont {Yang}}, \bibinfo {author}
  {\bibfnamefont {S.}~\bibnamefont {Pizzini}}, \bibinfo {author} {\bibfnamefont
  {D.}~\bibnamefont {de~Souza~Chaves}}, \bibinfo {author} {\bibfnamefont
  {A.}~\bibnamefont {Locatelli}}, \bibinfo {author} {\bibfnamefont {T.~O.}\
  \bibnamefont {Mente{\c{s}}}}, \bibinfo {author} {\bibfnamefont
  {A.}~\bibnamefont {Sala}}, \bibinfo {author} {\bibfnamefont {L.~D.}\
  \bibnamefont {Buda-Prejbeanu}}, \bibinfo {author} {\bibfnamefont
  {O.}~\bibnamefont {Klein}}, \emph {et~al.},\ }\bibfield  {title} {\bibinfo
  {title} {Room-temperature chiral magnetic skyrmions in ultrathin magnetic
  nanostructures},\ }\href@noop {} {\bibfield  {journal} {\bibinfo  {journal}
  {Nature nanotechnology}\ }\textbf {\bibinfo {volume} {11}},\ \bibinfo {pages}
  {449} (\bibinfo {year} {2016})}\BibitemShut {NoStop}%
\bibitem [{\citenamefont {M\"{u}hlbauer}\ \emph {et~al.}(2009)\citenamefont
  {M\"{u}hlbauer}, \citenamefont {Binz}, \citenamefont {Jonietz}, \citenamefont
  {Pfleiderer}, \citenamefont {Rosch}, \citenamefont {Neubauer}, \citenamefont
  {Georgii},\ and\ \citenamefont {B\"oni}}]{muh}%
  \BibitemOpen
  \bibfield  {author} {\bibinfo {author} {\bibfnamefont {S.}~\bibnamefont
  {M\"{u}hlbauer}}, \bibinfo {author} {\bibfnamefont {B.}~\bibnamefont {Binz}},
  \bibinfo {author} {\bibfnamefont {F.}~\bibnamefont {Jonietz}}, \bibinfo
  {author} {\bibfnamefont {C.}~\bibnamefont {Pfleiderer}}, \bibinfo {author}
  {\bibfnamefont {A.}~\bibnamefont {Rosch}}, \bibinfo {author} {\bibfnamefont
  {A.}~\bibnamefont {Neubauer}}, \bibinfo {author} {\bibfnamefont
  {R.}~\bibnamefont {Georgii}},\ and\ \bibinfo {author} {\bibfnamefont
  {P.}~\bibnamefont {B\"oni}},\ }\bibfield  {title} {\bibinfo {title} {Skyrmion
  lattice in a chiral magnet},\ }\href@noop {} {\bibfield  {journal} {\bibinfo
  {journal} {Science}\ }\textbf {\bibinfo {volume} {323}},\ \bibinfo {pages}
  {915} (\bibinfo {year} {2009})}\BibitemShut {NoStop}%
\bibitem [{\citenamefont {Dai}\ \emph {et~al.}(2023)\citenamefont {Dai},
  \citenamefont {Wu}, \citenamefont {Razavi}, \citenamefont {Xu}, \citenamefont
  {He}, \citenamefont {Shu}, \citenamefont {Jackson}, \citenamefont {Mahfouzi},
  \citenamefont {Huang}, \citenamefont {Pan} \emph {et~al.}}]{dai2023electric}%
  \BibitemOpen
  \bibfield  {author} {\bibinfo {author} {\bibfnamefont {B.}~\bibnamefont
  {Dai}}, \bibinfo {author} {\bibfnamefont {D.}~\bibnamefont {Wu}}, \bibinfo
  {author} {\bibfnamefont {S.~A.}\ \bibnamefont {Razavi}}, \bibinfo {author}
  {\bibfnamefont {S.}~\bibnamefont {Xu}}, \bibinfo {author} {\bibfnamefont
  {H.}~\bibnamefont {He}}, \bibinfo {author} {\bibfnamefont {Q.}~\bibnamefont
  {Shu}}, \bibinfo {author} {\bibfnamefont {M.}~\bibnamefont {Jackson}},
  \bibinfo {author} {\bibfnamefont {F.}~\bibnamefont {Mahfouzi}}, \bibinfo
  {author} {\bibfnamefont {H.}~\bibnamefont {Huang}}, \bibinfo {author}
  {\bibfnamefont {Q.}~\bibnamefont {Pan}}, \emph {et~al.},\ }\bibfield  {title}
  {\bibinfo {title} {Electric field manipulation of spin chirality and skyrmion
  dynamic},\ }\href@noop {} {\bibfield  {journal} {\bibinfo  {journal} {Science
  Advances}\ }\textbf {\bibinfo {volume} {9}},\ \bibinfo {pages} {eade6836}
  (\bibinfo {year} {2023})}\BibitemShut {NoStop}%
\bibitem [{\citenamefont {Kuznetsov}\ \emph {et~al.}(2023)\citenamefont
  {Kuznetsov}, \citenamefont {Mukhamatchin},\ and\ \citenamefont
  {Fraerman}}]{kuznetsov2023effective}%
  \BibitemOpen
  \bibfield  {author} {\bibinfo {author} {\bibfnamefont {M.}~\bibnamefont
  {Kuznetsov}}, \bibinfo {author} {\bibfnamefont {K.}~\bibnamefont
  {Mukhamatchin}},\ and\ \bibinfo {author} {\bibfnamefont {A.}~\bibnamefont
  {Fraerman}},\ }\bibfield  {title} {\bibinfo {title} {Effective interfacial
  dzyaloshinskii-moriya interaction and skyrmion stabilization in
  ferromagnet/paramagnet and ferromagnet/superconductor hybrid systems},\
  }\href@noop {} {\bibfield  {journal} {\bibinfo  {journal} {Physical Review
  B}\ }\textbf {\bibinfo {volume} {107}},\ \bibinfo {pages} {184428} (\bibinfo
  {year} {2023})}\BibitemShut {NoStop}%
\bibitem [{\citenamefont {Ba}\ \emph {et~al.}(2021)\citenamefont {Ba},
  \citenamefont {Zhuang}, \citenamefont {Zhang}, \citenamefont {Wang},
  \citenamefont {Gao}, \citenamefont {Zhou}, \citenamefont {Chen},
  \citenamefont {Sun}, \citenamefont {Liu}, \citenamefont {Chai}, \citenamefont
  {Ma}, \citenamefont {Zhang}, \citenamefont {Tian}, \citenamefont {Du},
  \citenamefont {Jiang}, \citenamefont {Nan}, \citenamefont {Hu},\ and\
  \citenamefont {Zhao}}]{Ba2021}%
  \BibitemOpen
  \bibfield  {author} {\bibinfo {author} {\bibfnamefont {Y.}~\bibnamefont
  {Ba}}, \bibinfo {author} {\bibfnamefont {S.}~\bibnamefont {Zhuang}}, \bibinfo
  {author} {\bibfnamefont {Y.}~\bibnamefont {Zhang}}, \bibinfo {author}
  {\bibfnamefont {Y.}~\bibnamefont {Wang}}, \bibinfo {author} {\bibfnamefont
  {Y.}~\bibnamefont {Gao}}, \bibinfo {author} {\bibfnamefont {H.}~\bibnamefont
  {Zhou}}, \bibinfo {author} {\bibfnamefont {M.}~\bibnamefont {Chen}}, \bibinfo
  {author} {\bibfnamefont {W.}~\bibnamefont {Sun}}, \bibinfo {author}
  {\bibfnamefont {Q.}~\bibnamefont {Liu}}, \bibinfo {author} {\bibfnamefont
  {G.}~\bibnamefont {Chai}}, \bibinfo {author} {\bibfnamefont {J.}~\bibnamefont
  {Ma}}, \bibinfo {author} {\bibfnamefont {Y.}~\bibnamefont {Zhang}}, \bibinfo
  {author} {\bibfnamefont {H.}~\bibnamefont {Tian}}, \bibinfo {author}
  {\bibfnamefont {H.}~\bibnamefont {Du}}, \bibinfo {author} {\bibfnamefont
  {W.}~\bibnamefont {Jiang}}, \bibinfo {author} {\bibfnamefont
  {C.}~\bibnamefont {Nan}}, \bibinfo {author} {\bibfnamefont {J.-M.}\
  \bibnamefont {Hu}},\ and\ \bibinfo {author} {\bibfnamefont {Y.}~\bibnamefont
  {Zhao}},\ }\bibfield  {title} {\bibinfo {title} {Electric-field control of
  skyrmions in multiferroic heterostructure via magnetoelectric coupling},\
  }\href {https://doi.org/10.1038/s41467-020-20528-y} {\bibfield  {journal}
  {\bibinfo  {journal} {Nature Communications}\ }\textbf {\bibinfo {volume}
  {12}},\ \bibinfo {pages} {322} (\bibinfo {year} {2021})}\BibitemShut
  {NoStop}%
\bibitem [{\citenamefont {Janoschek}\ \emph {et~al.}(2013)\citenamefont
  {Janoschek}, \citenamefont {Garst}, \citenamefont {Bauer}, \citenamefont
  {Krautscheid}, \citenamefont {Georgii}, \citenamefont {Boeni},\ and\
  \citenamefont {Pfleiderer}}]{janoschek2013fluctuation}%
  \BibitemOpen
  \bibfield  {author} {\bibinfo {author} {\bibfnamefont {M.}~\bibnamefont
  {Janoschek}}, \bibinfo {author} {\bibfnamefont {M.}~\bibnamefont {Garst}},
  \bibinfo {author} {\bibfnamefont {A.}~\bibnamefont {Bauer}}, \bibinfo
  {author} {\bibfnamefont {P.}~\bibnamefont {Krautscheid}}, \bibinfo {author}
  {\bibfnamefont {R.}~\bibnamefont {Georgii}}, \bibinfo {author} {\bibfnamefont
  {P.}~\bibnamefont {Boeni}},\ and\ \bibinfo {author} {\bibfnamefont
  {C.}~\bibnamefont {Pfleiderer}},\ }\bibfield  {title} {\bibinfo {title}
  {Fluctuation-induced first-order phase transition in dzyaloshinskii-moriya
  helimagnets},\ }\href@noop {} {\bibfield  {journal} {\bibinfo  {journal}
  {Physical Review B}\ }\textbf {\bibinfo {volume} {87}},\ \bibinfo {pages}
  {134407} (\bibinfo {year} {2013})}\BibitemShut {NoStop}%
\bibitem [{\citenamefont {Mishra}\ \emph {et~al.}(2023)\citenamefont {Mishra},
  \citenamefont {Samatham}, \citenamefont {Telling}, \citenamefont {Hillier},
  \citenamefont {Lees}, \citenamefont {Suresh},\ and\ \citenamefont
  {Ganesan}}]{mishra2023quantum}%
  \BibitemOpen
  \bibfield  {author} {\bibinfo {author} {\bibfnamefont {A.~K.}\ \bibnamefont
  {Mishra}}, \bibinfo {author} {\bibfnamefont {S.~S.}\ \bibnamefont
  {Samatham}}, \bibinfo {author} {\bibfnamefont {M.~T.}\ \bibnamefont
  {Telling}}, \bibinfo {author} {\bibfnamefont {A.}~\bibnamefont {Hillier}},
  \bibinfo {author} {\bibfnamefont {M.~R.}\ \bibnamefont {Lees}}, \bibinfo
  {author} {\bibfnamefont {K.}~\bibnamefont {Suresh}},\ and\ \bibinfo {author}
  {\bibfnamefont {V.}~\bibnamefont {Ganesan}},\ }\bibfield  {title} {\bibinfo
  {title} {Quantum griffiths phase in disordered mn 1- x fe x si},\ }\href@noop
  {} {\bibfield  {journal} {\bibinfo  {journal} {Physical Review B}\ }\textbf
  {\bibinfo {volume} {107}},\ \bibinfo {pages} {L100405} (\bibinfo {year}
  {2023})}\BibitemShut {NoStop}%
\bibitem [{\citenamefont {Nagaosa}\ and\ \citenamefont
  {Tokura}(2013)}]{nagaosa2013topological}%
  \BibitemOpen
  \bibfield  {author} {\bibinfo {author} {\bibfnamefont {N.}~\bibnamefont
  {Nagaosa}}\ and\ \bibinfo {author} {\bibfnamefont {Y.}~\bibnamefont
  {Tokura}},\ }\bibfield  {title} {\bibinfo {title} {Topological properties and
  dynamics of magnetic skyrmions},\ }\href@noop {} {\bibfield  {journal}
  {\bibinfo  {journal} {Nature nanotechnology}\ }\textbf {\bibinfo {volume}
  {8}},\ \bibinfo {pages} {899} (\bibinfo {year} {2013})}\BibitemShut {NoStop}%
\bibitem [{\citenamefont {Laref}\ \emph {et~al.}(2020)\citenamefont {Laref},
  \citenamefont {Kim},\ and\ \citenamefont {Manchon}}]{PhysRevB.102.060402}%
  \BibitemOpen
  \bibfield  {author} {\bibinfo {author} {\bibfnamefont {S.}~\bibnamefont
  {Laref}}, \bibinfo {author} {\bibfnamefont {K.-W.}\ \bibnamefont {Kim}},\
  and\ \bibinfo {author} {\bibfnamefont {A.}~\bibnamefont {Manchon}},\
  }\bibfield  {title} {\bibinfo {title} {Elusive dzyaloshinskii-moriya
  interaction in monolayer \({\mathrm{fe}}_{3}{\mathrm{gete}}_{2}\)},\ }\href
  {https://doi.org/10.1103/PhysRevB.102.060402} {\bibfield  {journal} {\bibinfo
   {journal} {Phys. Rev. B}\ }\textbf {\bibinfo {volume} {102}},\ \bibinfo
  {pages} {060402} (\bibinfo {year} {2020})}\BibitemShut {NoStop}%
\bibitem [{\citenamefont {Xu}\ \emph {et~al.}(2022)\citenamefont {Xu},
  \citenamefont {Li}, \citenamefont {Chen}, \citenamefont {Zhang},
  \citenamefont {Xiang},\ and\ \citenamefont
  {Bellaiche}}]{https://doi.org/10.1002/adma.202107779}%
  \BibitemOpen
  \bibfield  {author} {\bibinfo {author} {\bibfnamefont {C.}~\bibnamefont
  {Xu}}, \bibinfo {author} {\bibfnamefont {X.}~\bibnamefont {Li}}, \bibinfo
  {author} {\bibfnamefont {P.}~\bibnamefont {Chen}}, \bibinfo {author}
  {\bibfnamefont {Y.}~\bibnamefont {Zhang}}, \bibinfo {author} {\bibfnamefont
  {H.}~\bibnamefont {Xiang}},\ and\ \bibinfo {author} {\bibfnamefont
  {L.}~\bibnamefont {Bellaiche}},\ }\bibfield  {title} {\bibinfo {title}
  {Assembling diverse skyrmionic phases in fe3gete2 monolayers},\ }\href
  {https://onlinelibrary.wiley.com/doi/abs/10.1002/adma.202107779} {\bibfield
  {journal} {\bibinfo  {journal} {Advanced Materials}\ }\textbf {\bibinfo
  {volume} {34}},\ \bibinfo {pages} {2107779} (\bibinfo {year}
  {2022})}\BibitemShut {NoStop}%
\bibitem [{\citenamefont {Janson}\ \emph {et~al.}(2014)\citenamefont {Janson},
  \citenamefont {Rousochatzakis}, \citenamefont {Tsirlin}, \citenamefont
  {Belesi}, \citenamefont {Leonov}, \citenamefont {R{\"o}{\ss}ler},
  \citenamefont {van~den Brink},\ and\ \citenamefont {Rosner}}]{janson2014}%
  \BibitemOpen
  \bibfield  {author} {\bibinfo {author} {\bibfnamefont {O.}~\bibnamefont
  {Janson}}, \bibinfo {author} {\bibfnamefont {I.}~\bibnamefont
  {Rousochatzakis}}, \bibinfo {author} {\bibfnamefont {A.~A.}\ \bibnamefont
  {Tsirlin}}, \bibinfo {author} {\bibfnamefont {M.}~\bibnamefont {Belesi}},
  \bibinfo {author} {\bibfnamefont {A.~A.}\ \bibnamefont {Leonov}}, \bibinfo
  {author} {\bibfnamefont {U.~K.}\ \bibnamefont {R{\"o}{\ss}ler}}, \bibinfo
  {author} {\bibfnamefont {J.}~\bibnamefont {van~den Brink}},\ and\ \bibinfo
  {author} {\bibfnamefont {H.}~\bibnamefont {Rosner}},\ }\bibfield  {title}
  {\bibinfo {title} {The quantum nature of skyrmions and half-skyrmions in
  \({\mathrm{cu}}_{2}{\mathrm{oseo}}_{3}\)},\ }\href
  {https://doi.org/10.1038/ncomms6376} {\bibfield  {journal} {\bibinfo
  {journal} {Nature Communications}\ }\textbf {\bibinfo {volume} {5}},\
  \bibinfo {pages} {5376} (\bibinfo {year} {2014})}\BibitemShut {NoStop}%
\bibitem [{\citenamefont {Takao}(1983)}]{TAKAO19831009}%
  \BibitemOpen
  \bibfield  {author} {\bibinfo {author} {\bibfnamefont {S.}~\bibnamefont
  {Takao}},\ }\bibfield  {title} {\bibinfo {title} {A study of magnetization
  distribution of submicron bubbles in sputtered ho-co thin films},\ }\href
  {https://doi.org/https://doi.org/10.1016/0304-8853(83)90772-2} {\bibfield
  {journal} {\bibinfo  {journal} {Journal of Magnetism and Magnetic Materials}\
  }\textbf {\bibinfo {volume} {31-34}},\ \bibinfo {pages} {1009} (\bibinfo
  {year} {1983})}\BibitemShut {NoStop}%
\bibitem [{\citenamefont {Garel}\ and\ \citenamefont
  {Doniach}(1982)}]{garel1982phase}%
  \BibitemOpen
  \bibfield  {author} {\bibinfo {author} {\bibfnamefont {T.}~\bibnamefont
  {Garel}}\ and\ \bibinfo {author} {\bibfnamefont {S.}~\bibnamefont
  {Doniach}},\ }\bibfield  {title} {\bibinfo {title} {Phase transitions with
  spontaneous modulation-the dipolar ising ferromagnet},\ }\href@noop {}
  {\bibfield  {journal} {\bibinfo  {journal} {Physical Review B}\ }\textbf
  {\bibinfo {volume} {26}},\ \bibinfo {pages} {325} (\bibinfo {year}
  {1982})}\BibitemShut {NoStop}%
\bibitem [{\citenamefont {Nii}\ \emph {et~al.}(2015)\citenamefont {Nii},
  \citenamefont {Nakajima}, \citenamefont {Kikkawa}, \citenamefont {Yamasaki},
  \citenamefont {Ohishi}, \citenamefont {Suzuki}, \citenamefont {Taguchi},
  \citenamefont {Arima}, \citenamefont {Tokura},\ and\ \citenamefont
  {Iwasa}}]{nii2015uniaxial}%
  \BibitemOpen
  \bibfield  {author} {\bibinfo {author} {\bibfnamefont {Y.}~\bibnamefont
  {Nii}}, \bibinfo {author} {\bibfnamefont {T.}~\bibnamefont {Nakajima}},
  \bibinfo {author} {\bibfnamefont {A.}~\bibnamefont {Kikkawa}}, \bibinfo
  {author} {\bibfnamefont {Y.}~\bibnamefont {Yamasaki}}, \bibinfo {author}
  {\bibfnamefont {K.}~\bibnamefont {Ohishi}}, \bibinfo {author} {\bibfnamefont
  {J.}~\bibnamefont {Suzuki}}, \bibinfo {author} {\bibfnamefont
  {Y.}~\bibnamefont {Taguchi}}, \bibinfo {author} {\bibfnamefont
  {T.}~\bibnamefont {Arima}}, \bibinfo {author} {\bibfnamefont
  {Y.}~\bibnamefont {Tokura}},\ and\ \bibinfo {author} {\bibfnamefont
  {Y.}~\bibnamefont {Iwasa}},\ }\bibfield  {title} {\bibinfo {title} {Uniaxial
  stress control of skyrmion phase},\ }\href@noop {} {\bibfield  {journal}
  {\bibinfo  {journal} {Nature communications}\ }\textbf {\bibinfo {volume}
  {6}},\ \bibinfo {pages} {8539} (\bibinfo {year} {2015})}\BibitemShut
  {NoStop}%
\bibitem [{\citenamefont {Amoroso}\ \emph {et~al.}(2020)\citenamefont
  {Amoroso}, \citenamefont {Barone},\ and\ \citenamefont
  {Picozzi}}]{Amoroso2020}%
  \BibitemOpen
  \bibfield  {author} {\bibinfo {author} {\bibfnamefont {D.}~\bibnamefont
  {Amoroso}}, \bibinfo {author} {\bibfnamefont {P.}~\bibnamefont {Barone}},\
  and\ \bibinfo {author} {\bibfnamefont {S.}~\bibnamefont {Picozzi}},\
  }\bibfield  {title} {\bibinfo {title} {Spontaneous skyrmionic lattice from
  anisotropic symmetric exchange in a ni-halide monolayer},\ }\href
  {https://doi.org/10.1038/s41467-020-19535-w} {\bibfield  {journal} {\bibinfo
  {journal} {Nature Communications}\ }\textbf {\bibinfo {volume} {11}},\
  \bibinfo {pages} {5784} (\bibinfo {year} {2020})}\BibitemShut {NoStop}%
\bibitem [{\citenamefont {Wang}\ \emph {et~al.}(2016)\citenamefont {Wang},
  \citenamefont {Zhang}, \citenamefont {Xu}, \citenamefont {Peng},
  \citenamefont {Ding}, \citenamefont {Wang}, \citenamefont {Hou},
  \citenamefont {Zhang}, \citenamefont {Li}, \citenamefont {Liu}, \citenamefont
  {Wang}, \citenamefont {Cai}, \citenamefont {Wang}, \citenamefont {Li},
  \citenamefont {Hu}, \citenamefont {Wu}, \citenamefont {Shen},\ and\
  \citenamefont {Zhang}}]{https://doi.org/10.1002/adma.201600889}%
  \BibitemOpen
  \bibfield  {author} {\bibinfo {author} {\bibfnamefont {W.}~\bibnamefont
  {Wang}}, \bibinfo {author} {\bibfnamefont {Y.}~\bibnamefont {Zhang}},
  \bibinfo {author} {\bibfnamefont {G.}~\bibnamefont {Xu}}, \bibinfo {author}
  {\bibfnamefont {L.}~\bibnamefont {Peng}}, \bibinfo {author} {\bibfnamefont
  {B.}~\bibnamefont {Ding}}, \bibinfo {author} {\bibfnamefont {Y.}~\bibnamefont
  {Wang}}, \bibinfo {author} {\bibfnamefont {Z.}~\bibnamefont {Hou}}, \bibinfo
  {author} {\bibfnamefont {X.}~\bibnamefont {Zhang}}, \bibinfo {author}
  {\bibfnamefont {X.}~\bibnamefont {Li}}, \bibinfo {author} {\bibfnamefont
  {E.}~\bibnamefont {Liu}}, \bibinfo {author} {\bibfnamefont {S.}~\bibnamefont
  {Wang}}, \bibinfo {author} {\bibfnamefont {J.}~\bibnamefont {Cai}}, \bibinfo
  {author} {\bibfnamefont {F.}~\bibnamefont {Wang}}, \bibinfo {author}
  {\bibfnamefont {J.}~\bibnamefont {Li}}, \bibinfo {author} {\bibfnamefont
  {F.}~\bibnamefont {Hu}}, \bibinfo {author} {\bibfnamefont {G.}~\bibnamefont
  {Wu}}, \bibinfo {author} {\bibfnamefont {B.}~\bibnamefont {Shen}},\ and\
  \bibinfo {author} {\bibfnamefont {X.-X.}\ \bibnamefont {Zhang}},\ }\bibfield
  {title} {\bibinfo {title} {A centrosymmetric hexagonal magnet with
  superstable biskyrmion magnetic nanodomains in a wide temperature range of
  100–340 k},\ }\href
  {https://doi.org/https://doi.org/10.1002/adma.201600889} {\bibfield
  {journal} {\bibinfo  {journal} {Advanced Materials}\ }\textbf {\bibinfo
  {volume} {28}},\ \bibinfo {pages} {6887} (\bibinfo {year}
  {2016})}\BibitemShut {NoStop}%
\bibitem [{\citenamefont {Khanh}\ \emph {et~al.}(2020)\citenamefont {Khanh},
  \citenamefont {Nakajima}, \citenamefont {Yu}, \citenamefont {Gao},
  \citenamefont {Shibata}, \citenamefont {Hirschberger}, \citenamefont
  {Yamasaki}, \citenamefont {Sagayama}, \citenamefont {Nakao}, \citenamefont
  {Peng}, \citenamefont {Nakajima}, \citenamefont {Takagi}, \citenamefont
  {Arima}, \citenamefont {Tokura},\ and\ \citenamefont {Seki}}]{Khanh2020}%
  \BibitemOpen
  \bibfield  {author} {\bibinfo {author} {\bibfnamefont {N.~D.}\ \bibnamefont
  {Khanh}}, \bibinfo {author} {\bibfnamefont {T.}~\bibnamefont {Nakajima}},
  \bibinfo {author} {\bibfnamefont {X.}~\bibnamefont {Yu}}, \bibinfo {author}
  {\bibfnamefont {S.}~\bibnamefont {Gao}}, \bibinfo {author} {\bibfnamefont
  {K.}~\bibnamefont {Shibata}}, \bibinfo {author} {\bibfnamefont
  {M.}~\bibnamefont {Hirschberger}}, \bibinfo {author} {\bibfnamefont
  {Y.}~\bibnamefont {Yamasaki}}, \bibinfo {author} {\bibfnamefont
  {H.}~\bibnamefont {Sagayama}}, \bibinfo {author} {\bibfnamefont
  {H.}~\bibnamefont {Nakao}}, \bibinfo {author} {\bibfnamefont
  {L.}~\bibnamefont {Peng}}, \bibinfo {author} {\bibfnamefont {K.}~\bibnamefont
  {Nakajima}}, \bibinfo {author} {\bibfnamefont {R.}~\bibnamefont {Takagi}},
  \bibinfo {author} {\bibfnamefont {T.-h.}\ \bibnamefont {Arima}}, \bibinfo
  {author} {\bibfnamefont {Y.}~\bibnamefont {Tokura}},\ and\ \bibinfo {author}
  {\bibfnamefont {S.}~\bibnamefont {Seki}},\ }\bibfield  {title} {\bibinfo
  {title} {Nanometric square skyrmion lattice in a centrosymmetric tetragonal
  magnet},\ }\href {https://doi.org/10.1038/s41565-020-0684-7} {\bibfield
  {journal} {\bibinfo  {journal} {Nature Nanotechnology}\ }\textbf {\bibinfo
  {volume} {15}},\ \bibinfo {pages} {444} (\bibinfo {year} {2020})}\BibitemShut
  {NoStop}%
\bibitem [{\citenamefont {Yu}\ \emph {et~al.}(2012)\citenamefont {Yu},
  \citenamefont {Mostovoy}, \citenamefont {Tokunaga}, \citenamefont {Zhang},
  \citenamefont {Kimoto}, \citenamefont {Matsui}, \citenamefont {Kaneko},
  \citenamefont {Nagaosa},\ and\ \citenamefont
  {Tokura}}]{doi:10.1073/pnas.1118496109}%
  \BibitemOpen
  \bibfield  {author} {\bibinfo {author} {\bibfnamefont {X.}~\bibnamefont
  {Yu}}, \bibinfo {author} {\bibfnamefont {M.}~\bibnamefont {Mostovoy}},
  \bibinfo {author} {\bibfnamefont {Y.}~\bibnamefont {Tokunaga}}, \bibinfo
  {author} {\bibfnamefont {W.}~\bibnamefont {Zhang}}, \bibinfo {author}
  {\bibfnamefont {K.}~\bibnamefont {Kimoto}}, \bibinfo {author} {\bibfnamefont
  {Y.}~\bibnamefont {Matsui}}, \bibinfo {author} {\bibfnamefont
  {Y.}~\bibnamefont {Kaneko}}, \bibinfo {author} {\bibfnamefont
  {N.}~\bibnamefont {Nagaosa}},\ and\ \bibinfo {author} {\bibfnamefont
  {Y.}~\bibnamefont {Tokura}},\ }\bibfield  {title} {\bibinfo {title} {Magnetic
  stripes and skyrmions with helicity reversals},\ }\href
  {https://doi.org/10.1073/pnas.1118496109} {\bibfield  {journal} {\bibinfo
  {journal} {Proceedings of the National Academy of Sciences}\ }\textbf
  {\bibinfo {volume} {109}},\ \bibinfo {pages} {8856} (\bibinfo {year}
  {2012})}\BibitemShut {NoStop}%
\bibitem [{\citenamefont {Lin}\ and\ \citenamefont
  {Batista}(2018)}]{PhysRevLett.120.077202}%
  \BibitemOpen
  \bibfield  {author} {\bibinfo {author} {\bibfnamefont {S.-Z.}\ \bibnamefont
  {Lin}}\ and\ \bibinfo {author} {\bibfnamefont {C.~D.}\ \bibnamefont
  {Batista}},\ }\bibfield  {title} {\bibinfo {title} {Face centered cubic and
  hexagonal close packed skyrmion crystals in centrosymmetric magnets},\ }\href
  {https://doi.org/10.1103/PhysRevLett.120.077202} {\bibfield  {journal}
  {\bibinfo  {journal} {Phys. Rev. Lett.}\ }\textbf {\bibinfo {volume} {120}},\
  \bibinfo {pages} {077202} (\bibinfo {year} {2018})}\BibitemShut {NoStop}%
\bibitem [{\citenamefont {\ifmmode~\check{S}\else \v{S}\fi{}abani}\ \emph
  {et~al.}(2020)\citenamefont {\ifmmode~\check{S}\else \v{S}\fi{}abani},
  \citenamefont {Bacaksiz},\ and\ \citenamefont {Milo\ifmmode \check{s}\else
  \v{s}\fi{}evi\ifmmode~\acute{c}\else \'{c}\fi{}}}]{PhysRevB.102.014457}%
  \BibitemOpen
  \bibfield  {author} {\bibinfo {author} {\bibfnamefont {D.}~\bibnamefont
  {\ifmmode~\check{S}\else \v{S}\fi{}abani}}, \bibinfo {author} {\bibfnamefont
  {C.}~\bibnamefont {Bacaksiz}},\ and\ \bibinfo {author} {\bibfnamefont
  {M.~V.}\ \bibnamefont {Milo\ifmmode \check{s}\else
  \v{s}\fi{}evi\ifmmode~\acute{c}\else \'{c}\fi{}}},\ }\bibfield  {title}
  {\bibinfo {title} {Ab initio methodology for magnetic exchange parameters:
  Generic four-state energy mapping onto a heisenberg spin hamiltonian},\
  }\href {https://doi.org/10.1103/PhysRevB.102.014457} {\bibfield  {journal}
  {\bibinfo  {journal} {Phys. Rev. B}\ }\textbf {\bibinfo {volume} {102}},\
  \bibinfo {pages} {014457} (\bibinfo {year} {2020})}\BibitemShut {NoStop}%
\bibitem [{\citenamefont {Xiang}\ \emph {et~al.}(2011)\citenamefont {Xiang},
  \citenamefont {Kan}, \citenamefont {Wei}, \citenamefont {Whangbo},\ and\
  \citenamefont {Gong}}]{PhysRevB.84.224429}%
  \BibitemOpen
  \bibfield  {author} {\bibinfo {author} {\bibfnamefont {H.~J.}\ \bibnamefont
  {Xiang}}, \bibinfo {author} {\bibfnamefont {E.~J.}\ \bibnamefont {Kan}},
  \bibinfo {author} {\bibfnamefont {S.-H.}\ \bibnamefont {Wei}}, \bibinfo
  {author} {\bibfnamefont {M.-H.}\ \bibnamefont {Whangbo}},\ and\ \bibinfo
  {author} {\bibfnamefont {X.~G.}\ \bibnamefont {Gong}},\ }\bibfield  {title}
  {\bibinfo {title} {Predicting the spin-lattice order of frustrated systems
  from first principles},\ }\href {https://doi.org/10.1103/PhysRevB.84.224429}
  {\bibfield  {journal} {\bibinfo  {journal} {Phys. Rev. B}\ }\textbf {\bibinfo
  {volume} {84}},\ \bibinfo {pages} {224429} (\bibinfo {year}
  {2011})}\BibitemShut {NoStop}%
\bibitem [{\citenamefont {Xiang}\ \emph {et~al.}(2013)\citenamefont {Xiang},
  \citenamefont {Lee}, \citenamefont {Koo}, \citenamefont {Gong},\ and\
  \citenamefont {Whangbo}}]{C2DT31662E}%
  \BibitemOpen
  \bibfield  {author} {\bibinfo {author} {\bibfnamefont {H.}~\bibnamefont
  {Xiang}}, \bibinfo {author} {\bibfnamefont {C.}~\bibnamefont {Lee}}, \bibinfo
  {author} {\bibfnamefont {H.-J.}\ \bibnamefont {Koo}}, \bibinfo {author}
  {\bibfnamefont {X.}~\bibnamefont {Gong}},\ and\ \bibinfo {author}
  {\bibfnamefont {M.-H.}\ \bibnamefont {Whangbo}},\ }\bibfield  {title}
  {\bibinfo {title} {Magnetic properties and energy-mapping analysis},\ }\href
  {https://doi.org/10.1039/C2DT31662E} {\bibfield  {journal} {\bibinfo
  {journal} {Dalton Trans.}\ }\textbf {\bibinfo {volume} {42}},\ \bibinfo
  {pages} {823} (\bibinfo {year} {2013})}\BibitemShut {NoStop}%
\bibitem [{\citenamefont {Hoffmann}\ and\ \citenamefont
  {Bl\"ugel}(2020)}]{PhysRevB.101.024418}%
  \BibitemOpen
  \bibfield  {author} {\bibinfo {author} {\bibfnamefont {M.}~\bibnamefont
  {Hoffmann}}\ and\ \bibinfo {author} {\bibfnamefont {S.}~\bibnamefont
  {Bl\"ugel}},\ }\bibfield  {title} {\bibinfo {title} {Systematic derivation of
  realistic spin models for beyond-heisenberg solids},\ }\href
  {https://doi.org/10.1103/PhysRevB.101.024418} {\bibfield  {journal} {\bibinfo
   {journal} {Phys. Rev. B}\ }\textbf {\bibinfo {volume} {101}},\ \bibinfo
  {pages} {024418} (\bibinfo {year} {2020})}\BibitemShut {NoStop}%
\bibitem [{\citenamefont {Haldar}\ \emph {et~al.}(2018)\citenamefont {Haldar},
  \citenamefont {von Malottki}, \citenamefont {Meyer}, \citenamefont
  {Bessarab},\ and\ \citenamefont {Heinze}}]{PhysRevB.98.060413}%
  \BibitemOpen
  \bibfield  {author} {\bibinfo {author} {\bibfnamefont {S.}~\bibnamefont
  {Haldar}}, \bibinfo {author} {\bibfnamefont {S.}~\bibnamefont {von
  Malottki}}, \bibinfo {author} {\bibfnamefont {S.}~\bibnamefont {Meyer}},
  \bibinfo {author} {\bibfnamefont {P.~F.}\ \bibnamefont {Bessarab}},\ and\
  \bibinfo {author} {\bibfnamefont {S.}~\bibnamefont {Heinze}},\ }\bibfield
  {title} {\bibinfo {title} {First-principles prediction of sub-10-nm skyrmions
  in pd/fe bilayers on rh(111)},\ }\href
  {https://doi.org/10.1103/PhysRevB.98.060413} {\bibfield  {journal} {\bibinfo
  {journal} {Phys. Rev. B}\ }\textbf {\bibinfo {volume} {98}},\ \bibinfo
  {pages} {060413} (\bibinfo {year} {2018})}\BibitemShut {NoStop}%
\bibitem [{\citenamefont {Gorkan}\ \emph {et~al.}(2023)\citenamefont {Gorkan},
  \citenamefont {Das}, \citenamefont {Kapeghian}, \citenamefont {Akram},
  \citenamefont {Barth}, \citenamefont {Tongay}, \citenamefont {Akturk},
  \citenamefont {Erten},\ and\ \citenamefont
  {Botana}}]{PhysRevMaterials.7.054006}%
  \BibitemOpen
  \bibfield  {author} {\bibinfo {author} {\bibfnamefont {T.}~\bibnamefont
  {Gorkan}}, \bibinfo {author} {\bibfnamefont {J.}~\bibnamefont {Das}},
  \bibinfo {author} {\bibfnamefont {J.}~\bibnamefont {Kapeghian}}, \bibinfo
  {author} {\bibfnamefont {M.}~\bibnamefont {Akram}}, \bibinfo {author}
  {\bibfnamefont {J.~V.}\ \bibnamefont {Barth}}, \bibinfo {author}
  {\bibfnamefont {S.}~\bibnamefont {Tongay}}, \bibinfo {author} {\bibfnamefont
  {E.}~\bibnamefont {Akturk}}, \bibinfo {author} {\bibfnamefont
  {O.}~\bibnamefont {Erten}},\ and\ \bibinfo {author} {\bibfnamefont {A.~S.}\
  \bibnamefont {Botana}},\ }\bibfield  {title} {\bibinfo {title} {Skyrmion
  formation in ni-based janus dihalide monolayers: Interplay between magnetic
  frustration and dzyaloshinskii-moriya interaction},\ }\href
  {https://doi.org/10.1103/PhysRevMaterials.7.054006} {\bibfield  {journal}
  {\bibinfo  {journal} {Phys. Rev. Mater.}\ }\textbf {\bibinfo {volume} {7}},\
  \bibinfo {pages} {054006} (\bibinfo {year} {2023})}\BibitemShut {NoStop}%
\bibitem [{\citenamefont {Das}\ \emph {et~al.}(2023)\citenamefont {Das},
  \citenamefont {Gorkan}, \citenamefont {Kapeghian}, \citenamefont {Akram},
  \citenamefont {Tongay}, \citenamefont {Akt{\"u}rk}, \citenamefont {Erten},
  \citenamefont {Botana} \emph {et~al.}}]{das2023skyrmion}%
  \BibitemOpen
  \bibfield  {author} {\bibinfo {author} {\bibfnamefont {J.}~\bibnamefont
  {Das}}, \bibinfo {author} {\bibfnamefont {T.}~\bibnamefont {Gorkan}},
  \bibinfo {author} {\bibfnamefont {J.}~\bibnamefont {Kapeghian}}, \bibinfo
  {author} {\bibfnamefont {M.}~\bibnamefont {Akram}}, \bibinfo {author}
  {\bibfnamefont {S.}~\bibnamefont {Tongay}}, \bibinfo {author} {\bibfnamefont
  {E.}~\bibnamefont {Akt{\"u}rk}}, \bibinfo {author} {\bibfnamefont
  {O.}~\bibnamefont {Erten}}, \bibinfo {author} {\bibfnamefont
  {A.}~\bibnamefont {Botana}}, \emph {et~al.},\ }\bibfield  {title} {\bibinfo
  {title} {Skyrmion formation in ni-based janus dihalide monolayers: Interplay
  between magnetic frustration and dzyaloshinskii-moriya interaction},\
  }\href@noop {} {\bibfield  {journal} {\bibinfo  {journal} {Bulletin of the
  American Physical Society}\ } (\bibinfo {year} {2023})}\BibitemShut {NoStop}%
\bibitem [{\citenamefont {Dou}\ \emph {et~al.}(2022)\citenamefont {Dou},
  \citenamefont {Du}, \citenamefont {He}, \citenamefont {Dai}, \citenamefont
  {Huang},\ and\ \citenamefont {Ma}}]{dou2022theoretical}%
  \BibitemOpen
  \bibfield  {author} {\bibinfo {author} {\bibfnamefont {K.}~\bibnamefont
  {Dou}}, \bibinfo {author} {\bibfnamefont {W.}~\bibnamefont {Du}}, \bibinfo
  {author} {\bibfnamefont {Z.}~\bibnamefont {He}}, \bibinfo {author}
  {\bibfnamefont {Y.}~\bibnamefont {Dai}}, \bibinfo {author} {\bibfnamefont
  {B.}~\bibnamefont {Huang}},\ and\ \bibinfo {author} {\bibfnamefont
  {Y.}~\bibnamefont {Ma}},\ }\bibfield  {title} {\bibinfo {title} {Theoretical
  prediction of antiferromagnetic skyrmion crystal in janus monolayer
  crsi2n2as2},\ }\href@noop {} {\bibfield  {journal} {\bibinfo  {journal} {ACS
  nano}\ }\textbf {\bibinfo {volume} {17}},\ \bibinfo {pages} {1144} (\bibinfo
  {year} {2022})}\BibitemShut {NoStop}%
\bibitem [{\citenamefont {Han}\ \emph {et~al.}(2023)\citenamefont {Han},
  \citenamefont {Ji}, \citenamefont {Wang}, \citenamefont {Li},\ and\
  \citenamefont {Zhang}}]{han2023strain}%
  \BibitemOpen
  \bibfield  {author} {\bibinfo {author} {\bibfnamefont {Y.-t.}\ \bibnamefont
  {Han}}, \bibinfo {author} {\bibfnamefont {W.-x.}\ \bibnamefont {Ji}},
  \bibinfo {author} {\bibfnamefont {P.-J.}\ \bibnamefont {Wang}}, \bibinfo
  {author} {\bibfnamefont {P.}~\bibnamefont {Li}},\ and\ \bibinfo {author}
  {\bibfnamefont {C.-W.}\ \bibnamefont {Zhang}},\ }\bibfield  {title} {\bibinfo
  {title} {Strain-tunable skyrmions in two-dimensional monolayer janus
  magnets},\ }\href@noop {} {\bibfield  {journal} {\bibinfo  {journal}
  {Nanoscale}\ }\textbf {\bibinfo {volume} {15}},\ \bibinfo {pages} {6830}
  (\bibinfo {year} {2023})}\BibitemShut {NoStop}%
\bibitem [{\citenamefont {Li}\ \emph {et~al.}(2023)\citenamefont {Li},
  \citenamefont {Yu}, \citenamefont {Liang}, \citenamefont {Ga},\ and\
  \citenamefont {Yang}}]{li2023topological}%
  \BibitemOpen
  \bibfield  {author} {\bibinfo {author} {\bibfnamefont {P.}~\bibnamefont
  {Li}}, \bibinfo {author} {\bibfnamefont {D.}~\bibnamefont {Yu}}, \bibinfo
  {author} {\bibfnamefont {J.}~\bibnamefont {Liang}}, \bibinfo {author}
  {\bibfnamefont {Y.}~\bibnamefont {Ga}},\ and\ \bibinfo {author}
  {\bibfnamefont {H.}~\bibnamefont {Yang}},\ }\bibfield  {title} {\bibinfo
  {title} {Topological spin textures in 1 t-phase janus magnets: Interplay
  between dzyaloshinskii-moriya interaction, magnetic frustration, and
  isotropic higher-order interactions},\ }\href@noop {} {\bibfield  {journal}
  {\bibinfo  {journal} {Physical Review B}\ }\textbf {\bibinfo {volume}
  {107}},\ \bibinfo {pages} {054408} (\bibinfo {year} {2023})}\BibitemShut
  {NoStop}%
\bibitem [{\citenamefont {Liang}\ \emph {et~al.}(2020)\citenamefont {Liang},
  \citenamefont {Wang}, \citenamefont {Du}, \citenamefont {Hallal},
  \citenamefont {Garcia}, \citenamefont {Chshiev}, \citenamefont {Fert},\ and\
  \citenamefont {Yang}}]{liang2020very}%
  \BibitemOpen
  \bibfield  {author} {\bibinfo {author} {\bibfnamefont {J.}~\bibnamefont
  {Liang}}, \bibinfo {author} {\bibfnamefont {W.}~\bibnamefont {Wang}},
  \bibinfo {author} {\bibfnamefont {H.}~\bibnamefont {Du}}, \bibinfo {author}
  {\bibfnamefont {A.}~\bibnamefont {Hallal}}, \bibinfo {author} {\bibfnamefont
  {K.}~\bibnamefont {Garcia}}, \bibinfo {author} {\bibfnamefont
  {M.}~\bibnamefont {Chshiev}}, \bibinfo {author} {\bibfnamefont
  {A.}~\bibnamefont {Fert}},\ and\ \bibinfo {author} {\bibfnamefont
  {H.}~\bibnamefont {Yang}},\ }\bibfield  {title} {\bibinfo {title} {Very large
  dzyaloshinskii-moriya interaction in two-dimensional janus manganese
  dichalcogenides and its application to realize skyrmion states},\ }\href@noop
  {} {\bibfield  {journal} {\bibinfo  {journal} {Physical Review B}\ }\textbf
  {\bibinfo {volume} {101}},\ \bibinfo {pages} {184401} (\bibinfo {year}
  {2020})}\BibitemShut {NoStop}%
\bibitem [{\citenamefont {Xu}\ \emph {et~al.}(2020)\citenamefont {Xu},
  \citenamefont {Feng}, \citenamefont {Prokhorenko}, \citenamefont {Nahas},
  \citenamefont {Xiang},\ and\ \citenamefont {Bellaiche}}]{xu2020topological}%
  \BibitemOpen
  \bibfield  {author} {\bibinfo {author} {\bibfnamefont {C.}~\bibnamefont
  {Xu}}, \bibinfo {author} {\bibfnamefont {J.}~\bibnamefont {Feng}}, \bibinfo
  {author} {\bibfnamefont {S.}~\bibnamefont {Prokhorenko}}, \bibinfo {author}
  {\bibfnamefont {Y.}~\bibnamefont {Nahas}}, \bibinfo {author} {\bibfnamefont
  {H.}~\bibnamefont {Xiang}},\ and\ \bibinfo {author} {\bibfnamefont
  {L.}~\bibnamefont {Bellaiche}},\ }\bibfield  {title} {\bibinfo {title}
  {Topological spin texture in janus monolayers of the chromium trihalides cr
  (i, x) 3},\ }\href@noop {} {\bibfield  {journal} {\bibinfo  {journal}
  {Physical Review B}\ }\textbf {\bibinfo {volume} {101}},\ \bibinfo {pages}
  {060404} (\bibinfo {year} {2020})}\BibitemShut {NoStop}%
\bibitem [{\citenamefont {Zhang}\ \emph {et~al.}(2020)\citenamefont {Zhang},
  \citenamefont {Xu}, \citenamefont {Chen}, \citenamefont {Nahas},
  \citenamefont {Prokhorenko},\ and\ \citenamefont
  {Bellaiche}}]{zhang2020emergence}%
  \BibitemOpen
  \bibfield  {author} {\bibinfo {author} {\bibfnamefont {Y.}~\bibnamefont
  {Zhang}}, \bibinfo {author} {\bibfnamefont {C.}~\bibnamefont {Xu}}, \bibinfo
  {author} {\bibfnamefont {P.}~\bibnamefont {Chen}}, \bibinfo {author}
  {\bibfnamefont {Y.}~\bibnamefont {Nahas}}, \bibinfo {author} {\bibfnamefont
  {S.}~\bibnamefont {Prokhorenko}},\ and\ \bibinfo {author} {\bibfnamefont
  {L.}~\bibnamefont {Bellaiche}},\ }\bibfield  {title} {\bibinfo {title}
  {Emergence of skyrmionium in a two-dimensional crge (se, te) 3 janus
  monolayer},\ }\href@noop {} {\bibfield  {journal} {\bibinfo  {journal}
  {Physical Review B}\ }\textbf {\bibinfo {volume} {102}},\ \bibinfo {pages}
  {241107} (\bibinfo {year} {2020})}\BibitemShut {NoStop}%
\bibitem [{\citenamefont {Yagmurcukardes}\ \emph {et~al.}(2020)\citenamefont
  {Yagmurcukardes}, \citenamefont {Qin}, \citenamefont {Ozen}, \citenamefont
  {Sayyad}, \citenamefont {Peeters}, \citenamefont {Tongay},\ and\
  \citenamefont {Sahin}}]{yagmurcukardes2020quantum}%
  \BibitemOpen
  \bibfield  {author} {\bibinfo {author} {\bibfnamefont {M.}~\bibnamefont
  {Yagmurcukardes}}, \bibinfo {author} {\bibfnamefont {Y.}~\bibnamefont {Qin}},
  \bibinfo {author} {\bibfnamefont {S.}~\bibnamefont {Ozen}}, \bibinfo {author}
  {\bibfnamefont {M.}~\bibnamefont {Sayyad}}, \bibinfo {author} {\bibfnamefont
  {F.~M.}\ \bibnamefont {Peeters}}, \bibinfo {author} {\bibfnamefont
  {S.}~\bibnamefont {Tongay}},\ and\ \bibinfo {author} {\bibfnamefont
  {H.}~\bibnamefont {Sahin}},\ }\bibfield  {title} {\bibinfo {title} {Quantum
  properties and applications of 2d janus crystals and their superlattices},\
  }\href@noop {} {\bibfield  {journal} {\bibinfo  {journal} {Applied Physics
  Reviews}\ }\textbf {\bibinfo {volume} {7}} (\bibinfo {year}
  {2020})}\BibitemShut {NoStop}%
\bibitem [{\citenamefont {Xiao}\ \emph {et~al.}(2024)\citenamefont {Xiao},
  \citenamefont {Guan}, \citenamefont {Feng},\ and\ \citenamefont
  {Song}}]{xiao2024strain}%
  \BibitemOpen
  \bibfield  {author} {\bibinfo {author} {\bibfnamefont {R.}~\bibnamefont
  {Xiao}}, \bibinfo {author} {\bibfnamefont {Z.}~\bibnamefont {Guan}}, \bibinfo
  {author} {\bibfnamefont {D.}~\bibnamefont {Feng}},\ and\ \bibinfo {author}
  {\bibfnamefont {C.}~\bibnamefont {Song}},\ }\bibfield  {title} {\bibinfo
  {title} {Strain-tunable ferromagnetism and skyrmions in two-dimensional janus
  cr2xyte6 (x, y= si, ge, sn, and x$\ne$ y) monolayers},\ }\href@noop {}
  {\bibfield  {journal} {\bibinfo  {journal} {Journal of Applied Physics}\
  }\textbf {\bibinfo {volume} {135}} (\bibinfo {year} {2024})}\BibitemShut
  {NoStop}%
\bibitem [{\citenamefont {Luo}\ \emph {et~al.}(2018)\citenamefont {Luo},
  \citenamefont {Song}, \citenamefont {Li}, \citenamefont {Zhang},
  \citenamefont {Hong}, \citenamefont {Yang}, \citenamefont {Zou},
  \citenamefont {Xu},\ and\ \citenamefont
  {You}}]{doi:10.1021/acs.nanolett.7b04722}%
  \BibitemOpen
  \bibfield  {author} {\bibinfo {author} {\bibfnamefont {S.}~\bibnamefont
  {Luo}}, \bibinfo {author} {\bibfnamefont {M.}~\bibnamefont {Song}}, \bibinfo
  {author} {\bibfnamefont {X.}~\bibnamefont {Li}}, \bibinfo {author}
  {\bibfnamefont {Y.}~\bibnamefont {Zhang}}, \bibinfo {author} {\bibfnamefont
  {J.}~\bibnamefont {Hong}}, \bibinfo {author} {\bibfnamefont {X.}~\bibnamefont
  {Yang}}, \bibinfo {author} {\bibfnamefont {X.}~\bibnamefont {Zou}}, \bibinfo
  {author} {\bibfnamefont {N.}~\bibnamefont {Xu}},\ and\ \bibinfo {author}
  {\bibfnamefont {L.}~\bibnamefont {You}},\ }\bibfield  {title} {\bibinfo
  {title} {Reconfigurable skyrmion logic gates},\ }\href
  {https://doi.org/10.1021/acs.nanolett.7b04722} {\bibfield  {journal}
  {\bibinfo  {journal} {Nano Letters}\ }\textbf {\bibinfo {volume} {18}},\
  \bibinfo {pages} {1180} (\bibinfo {year} {2018})},\ \bibinfo {note} {pMID:
  29350935}\BibitemShut {NoStop}%
\bibitem [{\citenamefont {Sun}\ \emph {et~al.}(2020)\citenamefont {Sun},
  \citenamefont {Wang}, \citenamefont {Li}, \citenamefont {Zhang},
  \citenamefont {Chen}, \citenamefont {Wang},\ and\ \citenamefont
  {Cheng}}]{Sun2020}%
  \BibitemOpen
  \bibfield  {author} {\bibinfo {author} {\bibfnamefont {W.}~\bibnamefont
  {Sun}}, \bibinfo {author} {\bibfnamefont {W.}~\bibnamefont {Wang}}, \bibinfo
  {author} {\bibfnamefont {H.}~\bibnamefont {Li}}, \bibinfo {author}
  {\bibfnamefont {G.}~\bibnamefont {Zhang}}, \bibinfo {author} {\bibfnamefont
  {D.}~\bibnamefont {Chen}}, \bibinfo {author} {\bibfnamefont {J.}~\bibnamefont
  {Wang}},\ and\ \bibinfo {author} {\bibfnamefont {Z.}~\bibnamefont {Cheng}},\
  }\bibfield  {title} {\bibinfo {title} {Controlling bimerons as skyrmion
  analogues by ferroelectric polarization in 2d van der waals multiferroic
  heterostructures},\ }\href {https://doi.org/10.1038/s41467-020-19779-6}
  {\bibfield  {journal} {\bibinfo  {journal} {Nature Communications}\ }\textbf
  {\bibinfo {volume} {11}},\ \bibinfo {pages} {5930} (\bibinfo {year}
  {2020})}\BibitemShut {NoStop}%
\bibitem [{\citenamefont {Sun}\ \emph {et~al.}(2021)\citenamefont {Sun},
  \citenamefont {Wang}, \citenamefont {Zang}, \citenamefont {Li}, \citenamefont
  {Zhang}, \citenamefont {Wang},\ and\ \citenamefont
  {Cheng}}]{https://doi.org/10.1002/adfm.202104452}%
  \BibitemOpen
  \bibfield  {author} {\bibinfo {author} {\bibfnamefont {W.}~\bibnamefont
  {Sun}}, \bibinfo {author} {\bibfnamefont {W.}~\bibnamefont {Wang}}, \bibinfo
  {author} {\bibfnamefont {J.}~\bibnamefont {Zang}}, \bibinfo {author}
  {\bibfnamefont {H.}~\bibnamefont {Li}}, \bibinfo {author} {\bibfnamefont
  {G.}~\bibnamefont {Zhang}}, \bibinfo {author} {\bibfnamefont
  {J.}~\bibnamefont {Wang}},\ and\ \bibinfo {author} {\bibfnamefont
  {Z.}~\bibnamefont {Cheng}},\ }\bibfield  {title} {\bibinfo {title}
  {Manipulation of magnetic skyrmion in a 2d van der waals heterostructure via
  both electric and magnetic fields},\ }\href
  {https://doi.org/https://doi.org/10.1002/adfm.202104452} {\bibfield
  {journal} {\bibinfo  {journal} {Advanced Functional Materials}\ }\textbf
  {\bibinfo {volume} {31}},\ \bibinfo {pages} {2104452} (\bibinfo {year}
  {2021})}\BibitemShut {NoStop}%
\bibitem [{\citenamefont {Sun}\ \emph {et~al.}(2023)\citenamefont {Sun},
  \citenamefont {Wang}, \citenamefont {Yang}, \citenamefont {Li}, \citenamefont
  {Li}, \citenamefont {Huang},\ and\ \citenamefont
  {Cheng}}]{PhysRevB.107.184439}%
  \BibitemOpen
  \bibfield  {author} {\bibinfo {author} {\bibfnamefont {W.}~\bibnamefont
  {Sun}}, \bibinfo {author} {\bibfnamefont {W.}~\bibnamefont {Wang}}, \bibinfo
  {author} {\bibfnamefont {C.}~\bibnamefont {Yang}}, \bibinfo {author}
  {\bibfnamefont {X.}~\bibnamefont {Li}}, \bibinfo {author} {\bibfnamefont
  {H.}~\bibnamefont {Li}}, \bibinfo {author} {\bibfnamefont {S.}~\bibnamefont
  {Huang}},\ and\ \bibinfo {author} {\bibfnamefont {Z.}~\bibnamefont {Cheng}},\
  }\bibfield  {title} {\bibinfo {title} {Quantized movement of magnetic
  skyrmions in moir\'e multiferroic heterostructures},\ }\href
  {https://doi.org/10.1103/PhysRevB.107.184439} {\bibfield  {journal} {\bibinfo
   {journal} {Phys. Rev. B}\ }\textbf {\bibinfo {volume} {107}},\ \bibinfo
  {pages} {184439} (\bibinfo {year} {2023})}\BibitemShut {NoStop}%
\bibitem [{\citenamefont {Huang}\ \emph {et~al.}(2024)\citenamefont {Huang},
  \citenamefont {Schwartz}, \citenamefont {Shao}, \citenamefont {Kovalev},\
  and\ \citenamefont {Tsymbal}}]{huang2024magnetic}%
  \BibitemOpen
  \bibfield  {author} {\bibinfo {author} {\bibfnamefont {K.}~\bibnamefont
  {Huang}}, \bibinfo {author} {\bibfnamefont {E.}~\bibnamefont {Schwartz}},
  \bibinfo {author} {\bibfnamefont {D.-F.}\ \bibnamefont {Shao}}, \bibinfo
  {author} {\bibfnamefont {A.~A.}\ \bibnamefont {Kovalev}},\ and\ \bibinfo
  {author} {\bibfnamefont {E.~Y.}\ \bibnamefont {Tsymbal}},\ }\bibfield
  {title} {\bibinfo {title} {Magnetic antiskyrmions in two-dimensional van der
  waals magnets engineered by layer stacking},\ }\href@noop {} {\bibfield
  {journal} {\bibinfo  {journal} {Physical Review B}\ }\textbf {\bibinfo
  {volume} {109}},\ \bibinfo {pages} {024426} (\bibinfo {year}
  {2024})}\BibitemShut {NoStop}%
\bibitem [{\citenamefont {Hu}\ \emph {et~al.}(2023)\citenamefont {Hu},
  \citenamefont {Jin}, \citenamefont {Zhong}, \citenamefont {Dai},
  \citenamefont {Tao}, \citenamefont {Zhang}, \citenamefont {Han},
  \citenamefont {Jiang},\ and\ \citenamefont
  {Zhou}}]{doi:10.1021/acs.chemmater.3c00172}%
  \BibitemOpen
  \bibfield  {author} {\bibinfo {author} {\bibfnamefont {X.}~\bibnamefont
  {Hu}}, \bibinfo {author} {\bibfnamefont {Z.}~\bibnamefont {Jin}}, \bibinfo
  {author} {\bibfnamefont {Y.}~\bibnamefont {Zhong}}, \bibinfo {author}
  {\bibfnamefont {J.}~\bibnamefont {Dai}}, \bibinfo {author} {\bibfnamefont
  {X.}~\bibnamefont {Tao}}, \bibinfo {author} {\bibfnamefont {X.}~\bibnamefont
  {Zhang}}, \bibinfo {author} {\bibfnamefont {J.}~\bibnamefont {Han}}, \bibinfo
  {author} {\bibfnamefont {S.}~\bibnamefont {Jiang}},\ and\ \bibinfo {author}
  {\bibfnamefont {L.}~\bibnamefont {Zhou}},\ }\bibfield  {title} {\bibinfo
  {title} {Epitaxial growth of two-dimensional magnetic lateral and vertical
  heterostructures},\ }\href {https://doi.org/10.1021/acs.chemmater.3c00172}
  {\bibfield  {journal} {\bibinfo  {journal} {Chemistry of Materials}\ }\textbf
  {\bibinfo {volume} {35}},\ \bibinfo {pages} {4220} (\bibinfo {year}
  {2023})},\ \Eprint
  {https://arxiv.org/abs/https://doi.org/10.1021/acs.chemmater.3c00172}
  {https://doi.org/10.1021/acs.chemmater.3c00172} \BibitemShut {NoStop}%
\bibitem [{\citenamefont {Shen}\ \emph {et~al.}(2023)\citenamefont {Shen},
  \citenamefont {Zhou},\ and\ \citenamefont {Shen}}]{shen2023programmable}%
  \BibitemOpen
  \bibfield  {author} {\bibinfo {author} {\bibfnamefont {L.}~\bibnamefont
  {Shen}}, \bibinfo {author} {\bibfnamefont {Y.}~\bibnamefont {Zhou}},\ and\
  \bibinfo {author} {\bibfnamefont {K.}~\bibnamefont {Shen}},\ }\bibfield
  {title} {\bibinfo {title} {Programmable skyrmion-based logic gates in a
  single nanotrack},\ }\href@noop {} {\bibfield  {journal} {\bibinfo  {journal}
  {Physical Review B}\ }\textbf {\bibinfo {volume} {107}},\ \bibinfo {pages}
  {054437} (\bibinfo {year} {2023})}\BibitemShut {NoStop}%
\bibitem [{\citenamefont {{\v{S}}abani}\ \emph {et~al.}(2020)\citenamefont
  {{\v{S}}abani}, \citenamefont {Bacaksiz},\ and\ \citenamefont
  {Milo{\v{s}}evi{\'c}}}]{vsabani2020ab}%
  \BibitemOpen
  \bibfield  {author} {\bibinfo {author} {\bibfnamefont {D.}~\bibnamefont
  {{\v{S}}abani}}, \bibinfo {author} {\bibfnamefont {C.}~\bibnamefont
  {Bacaksiz}},\ and\ \bibinfo {author} {\bibfnamefont {M.}~\bibnamefont
  {Milo{\v{s}}evi{\'c}}},\ }\bibfield  {title} {\bibinfo {title} {Ab initio
  methodology for magnetic exchange parameters: Generic four-state energy
  mapping onto a heisenberg spin hamiltonian},\ }\href@noop {} {\bibfield
  {journal} {\bibinfo  {journal} {Physical Review B}\ }\textbf {\bibinfo
  {volume} {102}},\ \bibinfo {pages} {014457} (\bibinfo {year}
  {2020})}\BibitemShut {NoStop}%
\bibitem [{\citenamefont {Kohn}\ and\ \citenamefont
  {Sham}(1965)}]{PhysRev.140.A1133}%
  \BibitemOpen
  \bibfield  {author} {\bibinfo {author} {\bibfnamefont {W.}~\bibnamefont
  {Kohn}}\ and\ \bibinfo {author} {\bibfnamefont {L.~J.}\ \bibnamefont
  {Sham}},\ }\bibfield  {title} {\bibinfo {title} {Self-consistent equations
  including exchange and correlation effects},\ }\href
  {https://link.aps.org/doi/10.1103/PhysRev.140.A1133} {\bibfield  {journal}
  {\bibinfo  {journal} {Phys. Rev.}\ }\textbf {\bibinfo {volume} {140}},\
  \bibinfo {pages} {A1133} (\bibinfo {year} {1965})}\BibitemShut {NoStop}%
\bibitem [{\citenamefont {Bl\"ochl}(1994)}]{PhysRevB.50.17953}%
  \BibitemOpen
  \bibfield  {author} {\bibinfo {author} {\bibfnamefont {P.~E.}\ \bibnamefont
  {Bl\"ochl}},\ }\bibfield  {title} {\bibinfo {title} {Projector augmented-wave
  method},\ }\href {https://link.aps.org/doi/10.1103/PhysRevB.50.17953}
  {\bibfield  {journal} {\bibinfo  {journal} {Phys. Rev. B}\ }\textbf {\bibinfo
  {volume} {50}},\ \bibinfo {pages} {17953} (\bibinfo {year}
  {1994})}\BibitemShut {NoStop}%
\bibitem [{\citenamefont {Perdew}\ \emph {et~al.}(1996)\citenamefont {Perdew},
  \citenamefont {Burke},\ and\ \citenamefont
  {Ernzerhof}}]{PhysRevLett.77.3865}%
  \BibitemOpen
  \bibfield  {author} {\bibinfo {author} {\bibfnamefont {J.~P.}\ \bibnamefont
  {Perdew}}, \bibinfo {author} {\bibfnamefont {K.}~\bibnamefont {Burke}},\ and\
  \bibinfo {author} {\bibfnamefont {M.}~\bibnamefont {Ernzerhof}},\ }\bibfield
  {title} {\bibinfo {title} {Generalized gradient approximation made simple},\
  }\href {https://doi.org/10.1103/PhysRevLett.77.3865} {\bibfield  {journal}
  {\bibinfo  {journal} {Phys. Rev. Lett.}\ }\textbf {\bibinfo {volume} {77}},\
  \bibinfo {pages} {3865} (\bibinfo {year} {1996})}\BibitemShut {NoStop}%
\bibitem [{\citenamefont {Kresse}\ and\ \citenamefont
  {Hafner}(1993)}]{PhysRevB.47.558}%
  \BibitemOpen
  \bibfield  {author} {\bibinfo {author} {\bibfnamefont {G.}~\bibnamefont
  {Kresse}}\ and\ \bibinfo {author} {\bibfnamefont {J.}~\bibnamefont
  {Hafner}},\ }\bibfield  {title} {\bibinfo {title} {Ab initio molecular
  dynamics for liquid metals},\ }\href
  {https://doi.org/10.1103/PhysRevB.47.558} {\bibfield  {journal} {\bibinfo
  {journal} {Phys. Rev. B}\ }\textbf {\bibinfo {volume} {47}},\ \bibinfo
  {pages} {558} (\bibinfo {year} {1993})}\BibitemShut {NoStop}%
\bibitem [{\citenamefont {Kresse}\ and\ \citenamefont
  {Furthm\"{u}ller}(1996)}]{KRESSE199615}%
  \BibitemOpen
  \bibfield  {author} {\bibinfo {author} {\bibfnamefont {G.}~\bibnamefont
  {Kresse}}\ and\ \bibinfo {author} {\bibfnamefont {J.}~\bibnamefont
  {Furthm\"{u}ller}},\ }\bibfield  {title} {\bibinfo {title} {Efficiency of
  ab-initio total energy calculations for metals and semiconductors using a
  plane-wave basis set},\ }\href
  {https://doi.org/https://doi.org/10.1016/0927-0256(96)00008-0} {\bibfield
  {journal} {\bibinfo  {journal} {Computational Materials Science}\ }\textbf
  {\bibinfo {volume} {6}},\ \bibinfo {pages} {15} (\bibinfo {year}
  {1996})}\BibitemShut {NoStop}%
\bibitem [{ESI()}]{ESI}%
  \BibitemOpen
  \href@noop {} {\bibinfo {title} {{Supplementary Information}}}\BibitemShut
  {NoStop}%
\bibitem [{\citenamefont {Dudarev}\ \emph {et~al.}(1998)\citenamefont
  {Dudarev}, \citenamefont {Botton}, \citenamefont {Savrasov}, \citenamefont
  {Humphreys},\ and\ \citenamefont {Sutton}}]{PhysRevB.57.1505}%
  \BibitemOpen
  \bibfield  {author} {\bibinfo {author} {\bibfnamefont {S.~L.}\ \bibnamefont
  {Dudarev}}, \bibinfo {author} {\bibfnamefont {G.~A.}\ \bibnamefont {Botton}},
  \bibinfo {author} {\bibfnamefont {S.~Y.}\ \bibnamefont {Savrasov}}, \bibinfo
  {author} {\bibfnamefont {C.~J.}\ \bibnamefont {Humphreys}},\ and\ \bibinfo
  {author} {\bibfnamefont {A.~P.}\ \bibnamefont {Sutton}},\ }\bibfield  {title}
  {\bibinfo {title} {Electron-energy-loss spectra and the structural stability
  of nickel oxide: An lsda+u study},\ }\href
  {https://doi.org/10.1103/PhysRevB.57.1505} {\bibfield  {journal} {\bibinfo
  {journal} {Phys. Rev. B}\ }\textbf {\bibinfo {volume} {57}},\ \bibinfo
  {pages} {1505} (\bibinfo {year} {1998})}\BibitemShut {NoStop}%
\bibitem [{\citenamefont {Baroni}\ \emph {et~al.}(2001)\citenamefont {Baroni},
  \citenamefont {de~Gironcoli}, \citenamefont {Dal~Corso},\ and\ \citenamefont
  {Giannozzi}}]{RevModPhys.73.515}%
  \BibitemOpen
  \bibfield  {author} {\bibinfo {author} {\bibfnamefont {S.}~\bibnamefont
  {Baroni}}, \bibinfo {author} {\bibfnamefont {S.}~\bibnamefont
  {de~Gironcoli}}, \bibinfo {author} {\bibfnamefont {A.}~\bibnamefont
  {Dal~Corso}},\ and\ \bibinfo {author} {\bibfnamefont {P.}~\bibnamefont
  {Giannozzi}},\ }\bibfield  {title} {\bibinfo {title} {Phonons and related
  crystal properties from density-functional perturbation theory},\ }\href
  {https://doi.org/10.1103/RevModPhys.73.515} {\bibfield  {journal} {\bibinfo
  {journal} {Rev. Mod. Phys.}\ }\textbf {\bibinfo {volume} {73}},\ \bibinfo
  {pages} {515} (\bibinfo {year} {2001})}\BibitemShut {NoStop}%
\bibitem [{\citenamefont {Togo}\ and\ \citenamefont
  {Tanaka}(2015)}]{TOGO20151}%
  \BibitemOpen
  \bibfield  {author} {\bibinfo {author} {\bibfnamefont {A.}~\bibnamefont
  {Togo}}\ and\ \bibinfo {author} {\bibfnamefont {I.}~\bibnamefont {Tanaka}},\
  }\bibfield  {title} {\bibinfo {title} {First principles phonon calculations
  in materials science},\ }\href
  {https://doi.org/https://doi.org/10.1016/j.scriptamat.2015.07.021} {\bibfield
   {journal} {\bibinfo  {journal} {Scripta Materialia}\ }\textbf {\bibinfo
  {volume} {108}},\ \bibinfo {pages} {1} (\bibinfo {year} {2015})}\BibitemShut
  {NoStop}%
\bibitem [{\citenamefont {Krukau}\ \emph {et~al.}(2006)\citenamefont {Krukau},
  \citenamefont {Vydrov}, \citenamefont {Izmaylov},\ and\ \citenamefont
  {Scuseria}}]{krukau2006influence}%
  \BibitemOpen
  \bibfield  {author} {\bibinfo {author} {\bibfnamefont {A.~V.}\ \bibnamefont
  {Krukau}}, \bibinfo {author} {\bibfnamefont {O.~A.}\ \bibnamefont {Vydrov}},
  \bibinfo {author} {\bibfnamefont {A.~F.}\ \bibnamefont {Izmaylov}},\ and\
  \bibinfo {author} {\bibfnamefont {G.~E.}\ \bibnamefont {Scuseria}},\
  }\bibfield  {title} {\bibinfo {title} {Influence of the exchange screening
  parameter on the performance of screened hybrid functionals},\ }\href@noop {}
  {\bibfield  {journal} {\bibinfo  {journal} {The Journal of chemical physics}\
  }\textbf {\bibinfo {volume} {125}} (\bibinfo {year} {2006})}\BibitemShut
  {NoStop}%
\bibitem [{\citenamefont {Heyd}\ \emph {et~al.}(2003)\citenamefont {Heyd},
  \citenamefont {Scuseria},\ and\ \citenamefont {Ernzerhof}}]{heyd2003hybrid}%
  \BibitemOpen
  \bibfield  {author} {\bibinfo {author} {\bibfnamefont {J.}~\bibnamefont
  {Heyd}}, \bibinfo {author} {\bibfnamefont {G.~E.}\ \bibnamefont {Scuseria}},\
  and\ \bibinfo {author} {\bibfnamefont {M.}~\bibnamefont {Ernzerhof}},\
  }\bibfield  {title} {\bibinfo {title} {Hybrid functionals based on a screened
  coulomb potential},\ }\href@noop {} {\bibfield  {journal} {\bibinfo
  {journal} {The Journal of chemical physics}\ }\textbf {\bibinfo {volume}
  {118}},\ \bibinfo {pages} {8207} (\bibinfo {year} {2003})}\BibitemShut
  {NoStop}%
\bibitem [{\citenamefont {Landau}\ and\ \citenamefont
  {Binder}(2014)}]{landau2014guide}%
  \BibitemOpen
  \bibfield  {author} {\bibinfo {author} {\bibfnamefont {D.~P.}\ \bibnamefont
  {Landau}}\ and\ \bibinfo {author} {\bibfnamefont {K.}~\bibnamefont
  {Binder}},\ }\bibfield  {title} {\bibinfo {title} {A guide to monte carlo
  simulations in statistical physics},\ }\href@noop {} {\bibfield  {journal}
  {\bibinfo  {journal} {A Guide to Monte Carlo Simulations in Statistical
  Physics}\ } (\bibinfo {year} {2014})}\BibitemShut {NoStop}%
\bibitem [{\citenamefont {Marsaglia}(1972)}]{marsaglia1972choosing}%
  \BibitemOpen
  \bibfield  {author} {\bibinfo {author} {\bibfnamefont {G.}~\bibnamefont
  {Marsaglia}},\ }\bibfield  {title} {\bibinfo {title} {Choosing a point from
  the surface of a sphere},\ }\href@noop {} {\bibfield  {journal} {\bibinfo
  {journal} {The Annals of Mathematical Statistics}\ }\textbf {\bibinfo
  {volume} {43}},\ \bibinfo {pages} {645} (\bibinfo {year} {1972})}\BibitemShut
  {NoStop}%
\end{thebibliography}%
\end{document}

\title{Interface-Induced Ferromagnetism  in lateral $\mathrm{NiBr_{2}}$ and $\mathrm{NiCl_{2}}$ Heterostructure.}
\maketitle

\begin{figure}[!htb]
\begin{center}
\includegraphics[width=\linewidth]{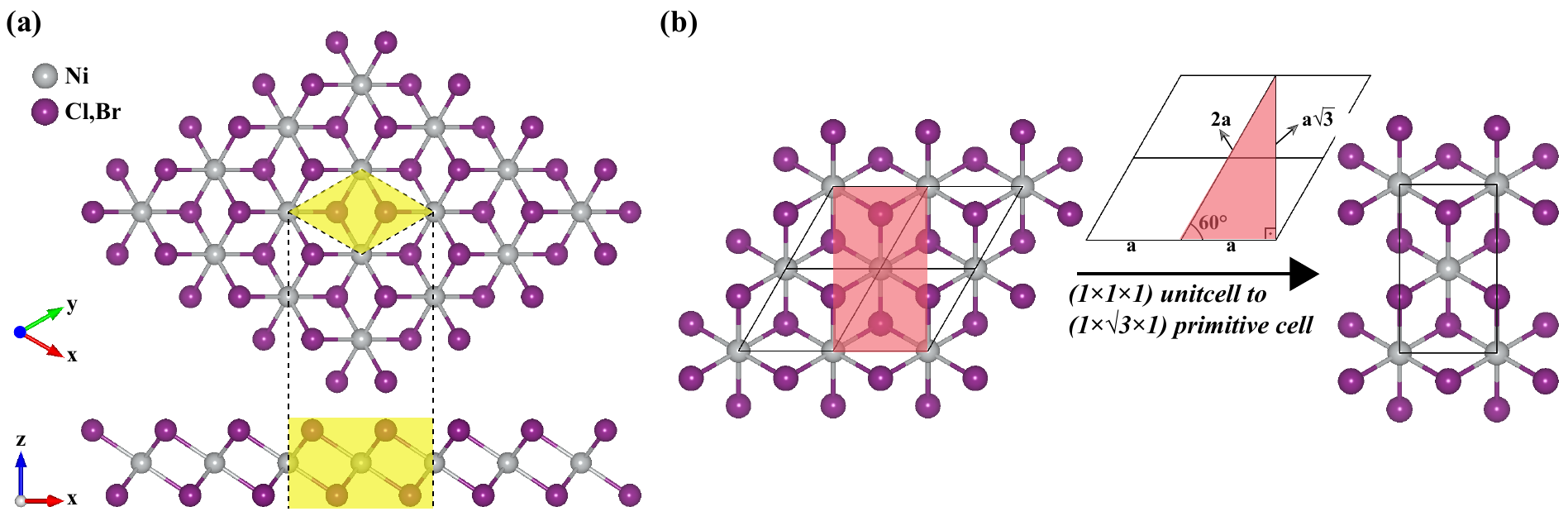}
\caption{\textbf{(a)} Top and side views of single-layer structures of $2\text{D NiX}_2$ ($\text{X}=\text{Cl, Br}$). The unit cell is indicated with dashed lines and yellow-shaded regions. \textbf{(b)} Visualization of the transition from the ($1 \times 1 \times 1$) unit cell to the ($1 \times \sqrt{3} \times 1$) cell to be used in heterostructure calculations.}
\label{S1}
\end{center}
\end{figure}

In Fig.~\ref{S1}, we illustrate the construction of orthogonal from a conventional unit cell of $\mathrm{NiBr_{2}}$ and $\mathrm{NiCl_{2}}$. 
  
\begin{figure}[!htb]
\begin{center}
\includegraphics[width=\linewidth]{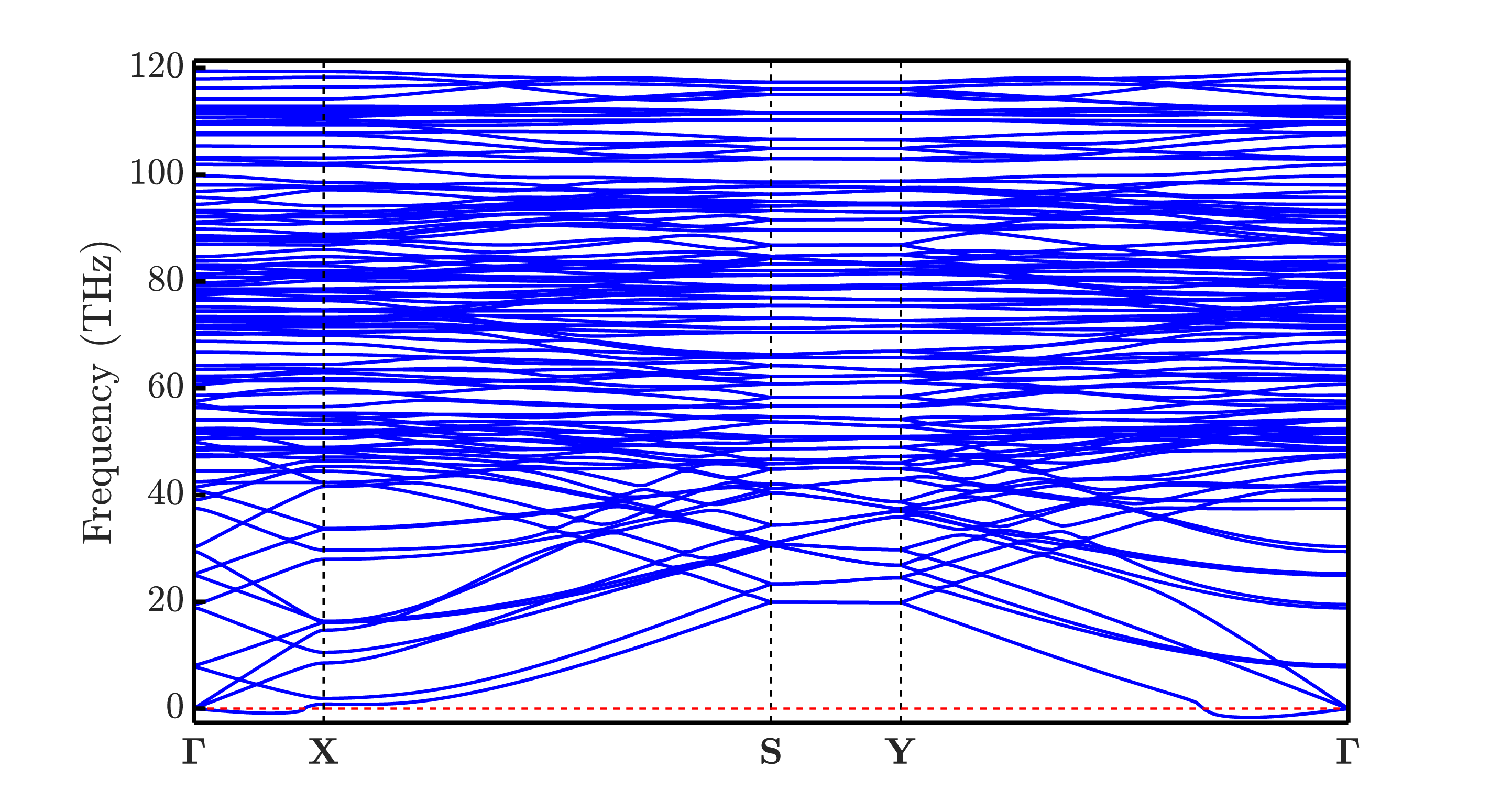}
\caption{ Phonon dispersion curves of $\mathrm{NiBr_{2}}$-$\mathrm{NiCl_{2}}$ heterostructure.  }
\label{S2}
\end{center}
\end{figure}

\begin{figure}[!htb]
\begin{center}
\includegraphics[width=\linewidth]{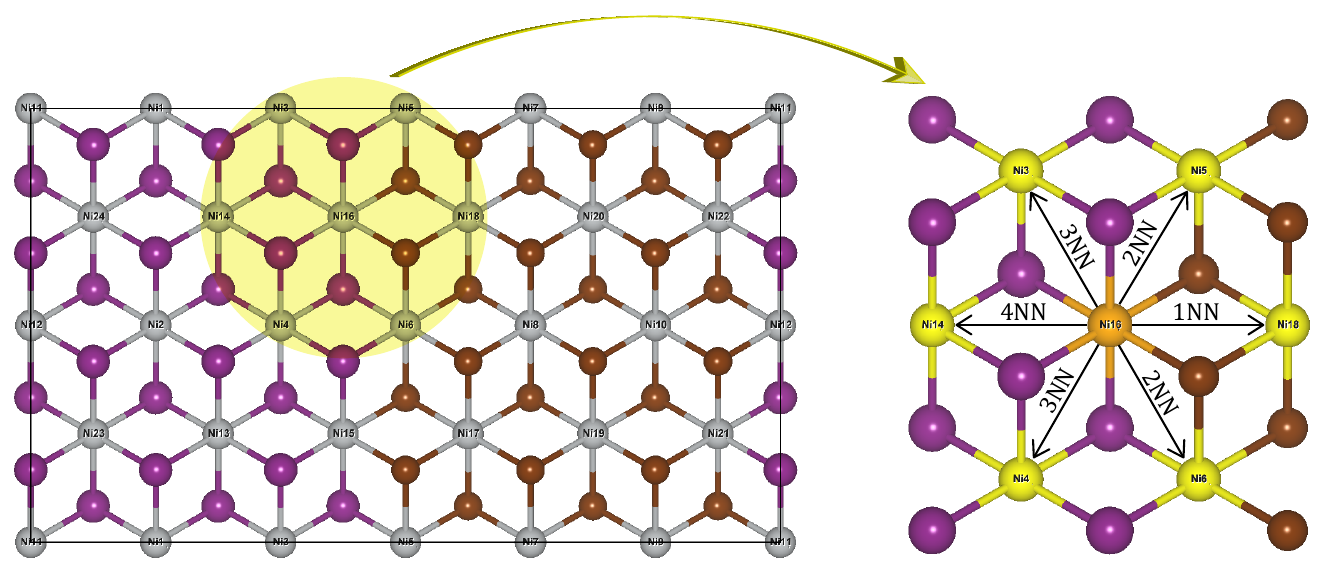}
\caption{{Left panel: $(6 \times 2 \sqrt{3} \times 1)$ supercell used in the calculations of four-state energy mapping in planar heterostructure. Right panel: Atom pairs selected for the four-state energy mapping method. The orange-colored atom represents the reference Ni atom, while the yellow-colored atoms represent neighboring magnetic Ni atoms to the reference atom.}}
\label{S3}
\end{center}
\end{figure}

For the $\mathrm{NiBr_{2}}$-$\mathrm{NiCl_{2}}$ interface, the lack of rotational symmetry requires to calculate the exchange-coupling tensor of some  individual nearest-neighbors. Fig.~\ref{S3} shows the neighbors considered in four-state energy mapping. Consequently, for the hetero-structure region, DFT calculations yield the following interaction tensors with the elements in units of meV regarding the neighboring dipole moments where the neighborhood indices appearing in the superscripts have been identified in Fig. \ref{S3}:

\begin{eqnarray}\nonumber\label{eq_S1}
	\mathbf{J}_{ij}^{16-18}&=&\left(
	\begin{array}{ccc}
		0.0057 & -0.00016 & 0.00075 \\
		-0.00016 & -0.0051 & -0.0021 \\
		0.00075 & -0.0021 & -0.0007 \\
	\end{array}
	\right),
	\quad
	\mathbf{J}_{ij}^{16-5}=\left(
	\begin{array}{ccc}
		0.043 & 0.068 & -0.11 \\
		0.068 & -0.049 & -0.057 \\
		-0.11 & -0.057 & 0.0056 \\
	\end{array}
	\right),
	\end{eqnarray}
\begin{eqnarray}\nonumber\label{eq_S2}
	\mathbf{J}_{ij}^{16-3}&=&\left(
	\begin{array}{ccc}
		0.077 & -0.11 & 0.15 \\
		-0.11 & -0.059 & -0.089 \\
		0.15 & -0.089 & -0.018 \\
	\end{array}
	\right),
		\mathbf{J}_{ij}^{16-14}=\left(
	\begin{array}{ccc}
		-0.13 & -0.0011 & 0.0011 \\
		-0.0011 & 0.13 & 0.17 \\
		0.0011 & 0.17 & -0.0061 \\
	\end{array}
	\right),
	\end{eqnarray}
with $\mathbf{J}_{ij}^{16-4}=\mathbf{J}_{ij}^{16-3},\mathbf{J}_{ij}^{16-6}=\mathbf{J}_{ij}^{16-5}$.
\begin{table}[h]
	%\begin{flushleft}
	\begin{center}
		\begin{threeparttable}
			\caption{Isotropic ferromagnetic exchange constants for the hetero-structure system.} \label{table1}
			\renewcommand{\arraystretch}{1.5}
			\begin{tabular}{ccccc}
				\thickhline
				Material & $J_{16-18}^{\mathrm{iso}}$ (meV) & $J_{16-5}^{\mathrm{iso}}$ (meV) & $J_{16-3}^{\mathrm{iso}}$ (meV) & $J_{16-14}^{\mathrm{iso}}$ (meV)\\
				\hline
				$\mathrm{NiBr_{2}+NiCl_{2}}$& -5.15 & -5.89 & -6.79  & -6.85 \\
				\thickhline \\
			\end{tabular}
			%\end{flushleft}
		\end{threeparttable}
	\end{center}
\end{table} 
\vspace{-1.0cm}

\begin{figure}[!ht]
\begin{center}

\includegraphics[width=0.75\linewidth]{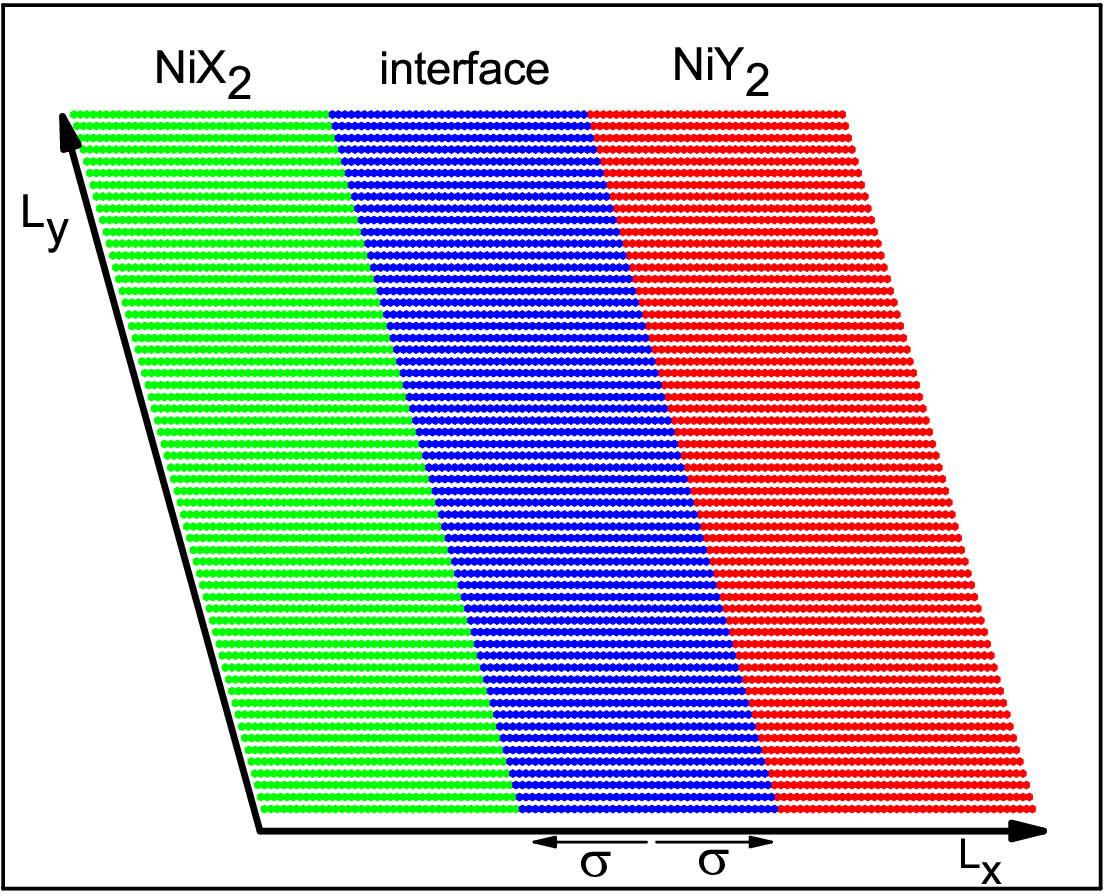}
\centering
\caption{Schematic representation of lateral $\mathrm{NiX_{2}+NiY_{2}(X=Br,Y=Cl)}$ hetero-structure system.}
\label{S4}
\end{center}

\end{figure}